\documentclass[12pt,dvips]{article}
\usepackage{here}
\usepackage{amsbsy,amssymb}
\newcommand{\be}[1]{\begin{equation}\label{#1}}
\newcommand{\ee}{\end{equation}}
\newcommand{\beq}[1]{\begin{eqnarray}\label{#1}}
\newcommand{\eeq}{\end{eqnarray}}
\newcommand{\ba}{\begin{array}}
\newcommand{\ea}{\end{array}}
\newcommand{\of}[1]{\left(#1\right)}
\newcommand{\off}[1]{\left[#1\right]}
\newcommand{\offf}[1]{\left\{#1\right\}}
\newcommand{\Tr}[1]{\mbox{Tr}\of{#1}}
\newcommand{\Trr}[1]{\mbox{Tr}\off{#1}}

\newcommand{\tr}[1]{\mbox{tr}\of{#1}}
\newcommand{\trr}[1]{\mbox{tr}\off{#1}}

\newcommand{\gb}[2]{\off{\!\off{#1,#2}\!}}
\newcommand{\bgb}[2]{\off{\!\!\off{#1,#2}\!\!}}
\newcommand{\bs}[1]{\boldsymbol{#1}}
\newcommand{\ms}[1]{\mbox{\small#1}}
\newcommand{\mf}[1]{\mbox{\footnotesize#1}}
\newcommand{\omb}{\bar\omega}
\newcommand{\vpb}{\bar\varpi}
\newcommand{\Par}{\par\vspace{0.2cm}\par}
\newcommand{\R}{{\cal R}}

\newcommand{\G}{{\cal G}}
\newcommand{\Q}{{\cal Q}}
\newcommand{\U}{{\cal U}}

\newcommand{\LL}{{\cal L}}
\newcommand{\gl}{\mathfrak{g}\mathfrak{l}}

\newcommand{\uu}{\mathfrak{u}}
\newcommand{\oo}{\mathfrak{o}}
\newcommand{\lf}{\mathfrak{l}}
\newcommand{\s}{\mathfrak{s}}

\newcommand{\qq}{\mathfrak{q}}
\renewcommand{\baselinestretch}{1.8}\tiny\normalsize
\begin{document}
\title{\renewcommand{\baselinestretch}{2.2}\tiny\normalsize
{\large\sf Post-Riemannian Merger of Yang-Mills
Interactions with Gravity}
\author{{\sc o. megged}\thanks{e-mail: megged@post.tau.ac.il}\\
{\small\rm School of Physics and Astronomy}\\
{\small\rm Tel-Aviv University, Tel-Aviv 69978, Israel.}}}
\date{}
\maketitle
\begin{abstract}
We show that a post-Riemannian spacetime can accommodate an internal
symmetry structure of the Yang-Mills prototype in such a way that the 
internal symmetry becomes an integral part of the spacetime itself.  
The construction encrusts the internal degrees of
freedom in spacetime in a manner that merges the gauging of
these degrees of freedom with the frame geometrical gauges of
spacetime. In particular, we prove that the three spacetime structural 
identities, which now become ``contaminated'' by internal degrees of
freedom, remain invariant with respect to internal gauge
transformations.
In a Weyl Cartan spacetime, the theory regains the original
form of Einstein's equations, in which gauge field sources on the 
r.h.s. determine on the l.h.s the geometry of spacetime and the fields
it induces. In the more general case we
identify new contributions of weak magnitude in the
interaction between the Yang-Mills field and gravity. The merger of
spacetime with internal degrees of freedom which we propose here is
not constrained by the usual Coleman-Mandula considerations. 
\end{abstract}
\newpage
%******************
\section{\sf Introduction}\label{s1}
%******************
In this paper we prove an equivalence theorem between, on the one
hand, Yang-Mills field, gauging an internal Abelian or
non-Abelian symmetry group $U$, {\em in the presence of Einstein
gravity\/}, and on the other hand, a regrouping in which Einstein's
gravity {\em and\/} the internal symmetry are replaced by an {\em
extended post-Riemannian spacetime theory\/} of the Weyl-Cartan type,
which includes 
torsion and Weyl's non-metricity. The theory thereby regains the
original structure of Einstein's equations, in which gauge
field sources on the r.h.s. determine on the l.h.s the geometry of
spacetime and the fields it induces.\Par
The construction {\em encrusts\/} the internal degrees of freedom in
spacetime in such a way as to {\em merge\/} the gauging of these
internal degrees of freedom with the frame geometrical gauges of
spacetime. Note that {\em unification\/} of
gravity with gauge theories, such as those of the Standard Model, has
up to now been achieved either through supersymmetry (as in
supergravity) or with new spatial dimensions (as in Kaluza-Klein
methodology). Note that putting together the Gravity and Yang-Mills
connection, i.e. producing the two from the same fabric - this does
occur in String Theory, where both come as {\em excitations of
strings\/}, either closed or open.\Par
This equivalence theorem is extended in a straightforward manner, so
that the resulting merged solution be made to correspond to more
general post-Riemannian configurations and their corresponding
spacetime theory. However, whereas the Weyl-Cartan regrouping may be
regarded as no more than a formal (but by no means trivial)
geometrical pasting of gauge symmetries, here the presence of Abelian
components in the internal gauge group gives rise to new
contributions of strength weaker than the gravitational magnitude in
the interaction between the Yang-Mills field and gravity.\Par   
Note that this {\em merger\/} of spacetime with internal degrees of
freedom is not constrained by the usual Coleman-Mandula
considerations. The physical states in the Hilbert space of the gauge
particles are unmodified and can anyhow only be presented in the
original input picture of the Yang-Mills field (and this only in
quasi-flat background, i.e. if it can be assumed that it occurs in the 
presence of weak gravity) as we have no particle Hilbert space
description containing information about the geometry of
spacetime.\Par
The structure and the content of this paper go as follows:
\begin{description}
\item{\bf{\em Section \ref{s2}}}\, is a pedagogical survey served to
fix the language, and to define the framework and tools we shall later
intensively utilize. A brief (but comprehensive) analysis of the
structure of spacetime with torsion and non-metricity is given, with an
emphasis on the gauge principle underlying the theory. 
Two auxiliary constructions which improve the
derivations have been included in this prelude:
\begin{enumerate}
\item
Replacing the conventional ``{\em reductive\/}'' formalism for
post-Riemannian geometry by a ``{\em constructive\/}''
approach. Explicitly, the reductive approach consists in
defining the connection and frames for an-holonomic
$GL\of{n,\mathbb{R}}$, then splitting the geometry into its Riemannian
piece with the Christoffel connection gauging the orthogonal subgroup,
the Weyl component gauging dilations, and the traceless non-metricity 
component gauging the shears. In the constructive
systematics, the connection and the frames are originally introduced
in the Riemannian framework, with the connection then undergoing a
{\em deformation\/}. The post-Riemannian structure then emerges as a
deformation of the original Riemannian structure. 
\item
Extensive use of absolute differentials.
\end{enumerate}
\item{\bf{\em section \ref{s3}}}\, presents a formalism in which
internal symmetry is naturally embedded {\em within\/} the non-metric
part of spacetime such that the internal symmetry becomes an 
integral part of the spacetime itself. Namely, the basespace
components of the gauge field of the embedded symmetry, {\em
dictated\/} to be uniquely of the Yang-Mills prototype, are unified
with those of the spacetime connection. Thereby, our merging theory
relies on a {\em single\/} connection 1-form; we shall elaborate on
its non-usual algebraic structure, and on the means by which the
metric sector remains protected against invasions from the internal
world. We shall explicitly re-derive the relevant spacetime
structural identities (now containing internal
degrees of freedom as well), and prove that they are invariant with
respect to internal gauge transformations.
\item{\bf{\em Section \ref{s4}\/}}\, deals with dynamics.
Starting from a minimal variant of
the Einstein-Hilbert action, suitable for a post-Riemannian spacetime,
we analyze in details the dynamical structure of the Weyl-Cartan
merger. In particular, we re-derive Einstein's equation in terms of
a fancy equality between the Einstein $\of{n-1}$-form and the
energy-momentum current composed of the Yang-Mills field strength. 
The latter object is shown to be traceless only in four dimensions, in
which case self dual solutions to the Yang-Mills equations are seen to 
correspond to {\em non-trivial gravitational emptiness\/}. Working in
a preferred gauge, we turn to study the more general case in
a qualitative manner. The gauge fixing process endows 
the gauge fields with a scaling function, by thus making them sensitive
to the non-metric microstructure of spacetime. We isolate those terms
in the action in which these fields interact with the
gravitational field, and find that the magnitude of these 
interactions is weaker than the gravitational magnitude. 
\item{\bf{\em Section \ref{s5}}}\, contains summarizing and closing
remarks. 
\end{description}\Par
We have also included in this work a {\bf{\em supplement}\/}
where we develop some integration formulas, and discuss some
of the topological aspects of the theory. Among other things, we
calculate the charge of the 1-st Chern class associated with the
merger, and 
formulate analytically two {\em non-Abelian\/} generalizations to
Stokes' theorem. The first of them concerns the Weyl covector (a
formula to which we also refer in the text); the second concerns the
coframe-torsion transvection field. 
\Par\Par\Par\noindent
{\em Some technical conventions\/}:\Par\noindent
%-------------
In what follows we shall extensively employ Cartan's differential
calculus, reinforced by the powerful concept of the {\em absolute
differential\/} \cite{LF} which is a generalized exterior
derivative: it operates as an ordinary exterior derivative on forms,
but when it is applied to objects such as vectors and tensors, it
generates (here $O\of{n}$-) covariant exterior derivatives. A rough
(and rather intuitive) definition will be given below.\Par 
Consider a tensor $p$-form $\bs{t}_{p}$ which we may schematically
write as $\bs{t}_{p}={t\off{\vartheta}}\off{\bs{e}}$. Here $t$ denotes its
components (or coordinates), $\off{\vartheta}$ the Grassmann basis,
and $\off{\bs{e}}$ the tensorial basis. The 1-st and 2-nd order
absolute differentials of $\bs{t}_{p}$ are given by:
\be{i3}\ba{rcl}
d\bs{t}_{p}&=&{d\of{t\off{\vartheta}}}\off{\bs{e}}+
\of{-1}^{p}t\off{\vartheta}\wedge{d}\off{\bs{e}},\\
dd\bs{t}_{p}&=&t\off{\vartheta}\wedge{dd}\off{\bs{e}};
\ea\ee
note that in our paradigm $d^{2}$ fails to annihilate on tensor
bases. Instead, as with covariant exterior derivatives, it will be
shown to generate the $O\of{n}$-curvature. But for scalar-valued
forms we have: $d\of{t\off{\vartheta}}=
dt\wedge\off{\vartheta}+td\off{\vartheta}$,\, and
clearly $d^{2}\of{t\off{\vartheta}}=0$.\Par   
Round brackets around a cluster of indices of an object, as in
$Q_{\alpha\of{\beta\gamma}}$, designate complete 
symmetrization; square brackets instead, as in $R_{\off{\alpha\beta}}$,
designate complete anti-symmetrization (no factorial terms are
included: $R_{\off{\alpha\beta}}:=R_{\alpha\beta}-R_{\beta\alpha}$
etc.). Bars added in between indices arranged in a cluster,  
such as in $\varpi_{\off{\alpha\left|\beta\right|\gamma}}$, tell which
indices are to be excluded in the symmetrization process (here
$\beta$). We shall often employ graded brackets, 
\be{i4}
\gb{\alpha}{\beta}:=\alpha\wedge\beta-
\of{-1}^{\of{\mf{deg}\alpha}\of{\mf{deg}\beta}}
\beta\wedge\alpha,
\ee
and only in few occasions refer to the ordinary commutator,
$\off{\alpha,\beta}=\alpha\wedge\beta-\beta\wedge\alpha$.\Par
The substratum in our model is an $n$ dimensional
differentiable manifold which we shall simply denote by $\Omega$ (the
events continuum). The signature it carries is $\of{p,q}$ 
$\of{p+q=n}$ but our formalism, in general, is not
sensitive to the signature. We shall 
sometimes refer to an arbitrary $m$-domain $\of{m\leq{n}}$, in
which case we shall employ the letter $\Sigma$ instead. Finally, we
use the notation 
${T\Omega}^{\otimes{k}}$ to denote the bundle of product space 
fibers, each of whose fibers is a space product of $k$ copies of the
tangent space (or the bundle of rank-$k$ tensors over $\Omega$).\Par  
%******************
\section{\sf A spacetime with torsion and non-metricity}\label{s2}
%******************
%----------------------
\subsection{\sf Employing the bundle of tangent frames by means
of absolute differentials}\label{s21}
%----------------------
The building blocks in the gauge theory of spacetime are the {\em frame
fields\/}. Until otherwise stated, holonomic frame elements attach a
Greek index, ${\bs{e}_{\alpha}}$, whereas non-holonomic frames
attach a Latin index, ${\bs{e}_{a}}$; 
note: the $\offf{\bs{e}_{a}}$ system is not a-priori
constrained to be rectilinear. The elements of the metric tensor in the
holonomic and the an-holonomic bases are given, respectively, by 
$g_{\alpha\beta}:=\bs{e}_{\alpha}\cdot\bs{e}_{\beta}$ and
$\bar{g}_{ab}:=\bs{e}_{a}\cdot\bs{e}_{b}$, by means of which the
scalar product is defined. The inverse metric tensor is assumed to
exist, and its elements with superscript indices are defined through
the relations
$g_{\alpha\beta}{g}^{\beta\gamma}=\delta_{\alpha}^{\gamma}$, and  
$\bar{g}_{ab}{\bar{g}}^{bc}=\delta^{c}_{a}$.\Par 
Substituting ${\bs{e}}^{\alpha}:=\bs{e}_{\beta}{g}^{\beta\alpha}$, and
${\bs{e}}^{a}:=\bs{e}_{b}\bar{g}^{ba}$ for {\em inverse frames\/},
leads to the orthonormality relations
$\bs{e}_{\alpha}\cdot{\bs{e}}^{\beta}=\delta_{\alpha}^{\beta}$,
and $\bs{e}_{a}\cdot{\bs{e}}^{b}=\delta_{a}^{b}$, whence
$\bs{e}_{\alpha}\cdot{\bs{e}}^{\alpha}=\bs{e}_{a}\cdot{\bs{e}}^{a}=n$.
Notice: the inverse frame shouldn't be confused with the concept of a
{\em coframe\/} discussed later in section \ref{s24}.
The {\em $n$-bein\/} $e_{\alpha}^{a}$ in the linear expansion 
$\bs{e}_{\alpha}=e_{\alpha}^{a}\bs{e}_{a}$, constitutes the
{\em coordinate vector\/} of $\bs{e}_{\alpha}$ in the basis
$\offf{\bs{e}_{a}}$ at $x$. Consequently, the
relation $g_{\alpha\beta}=e_{\alpha}^{a}e_{\beta}^{b}\bar{g}_{ab}$ is
frequently interpreted as a dot product between two coordinate vectors
with $x$-dependent mediating metric tensor $\bar{\bs{g}}$. From
$\bs{e}_{\alpha}=e_{\alpha}^{a}\bs{e}_{a}=:
e_{\alpha}^{a}\bar{e}^{\beta}_{a}\bs{e}_{\beta}$
one deduces that
$e_{\alpha}^{a}\bar{e}^{\beta}_{a}=\delta_{\alpha}^{\beta}$, and
similarly, from $\bs{e}_{a}=\bar{e}_{a}^{\alpha}\bs{e}_{\alpha}=
\bar{e}_{a}^{\alpha}e_{\alpha}^{b}\bs{e}_{b}$
one has: $e_{\alpha}^{a}\bar{e}^{\alpha}_{b}=\delta^{a}_{b}$.\Par 
A {\em metric-compatible\/} connection $\omega$ can locally be defined
through the {\em absolute differential\/} of a frame
field.\footnote{The concept of the absolute differential is rigorously
presented in \cite{LF} where it is also extensively used.} In a
holonomic basis,
\be{5}
\lim_{\Delta{x}\rightarrow{0}}\off{
\left.\bs{e}_{\alpha}\of{x+\Delta{x}}\right|_{x}-
\left.\bs{e}_{\alpha}\of{x}\right|_{x}}\,=:\,
-\omega^{\beta}_{\;\:\alpha}\of{x}
\left.\bs{e}_{\beta}\of{x}\right|_{x};
\ee
namely, we expand the difference between the value of a given frame
field at $x+\Delta{x}$ (basis vector for the tangent space at
$x+\Delta{x}$), displaced in a ``parallel'' manner to the tangent space at
$x$ where we take its measure, and the value of that same frame field
at $x$ (basis vector for the tangent space at $x$), in terms of
the frame system at $x$, as the distance $\Delta{x}$ on $\Omega$
between the two points (associated with these two tangent spaces)
approaches zero.\footnote{If applied to {\em functions\/} on $\Omega$,
the limiting procedure in the l.h.s. of (\ref{5}) generates ordinary
differentials.}\Par
Eq. (\ref{5}) may also take the abbreviated 
form:\footnote{From $d\of{\bs{e}_{\alpha}\cdot\bs{e}^{\beta}}=0$, 
the absolute differential of an inverse frame must come
with $+\omega$, %instead of with $-\omega$, 
$d\bs{e}^{\alpha}=\omega_{\;\:\beta}^{\alpha}\bs{e}^{\beta}$.} 
\be{5a}
\ba{rcl}
d\bs{e}_{\alpha}=-\omega_{\;\:\alpha}^{\beta}\bs{e}_{\beta}
&\stackrel{\ms{def}}{\Rightarrow}&
\of{D_{\omega}\bs{e}}_{\alpha}=0;
\ea
\ee
the frames are then said to be covariant-free, or {\em
parallel\/}. But: the value of the vector 1-form $d\bs{e}_{\alpha}$ at
$x$ depends on how the basis vector $\bs{e}_{\alpha}\of{x+\Delta{x}}$
was transported from the tangent space at $x+\Delta{x}$ to the tangent
space at $x$. The displacement method is reflected in the
form of the expansion coefficients, and it is only in this sense that
the frames are regarded as parallel. In an obvious manner, in a
non-holonomic basis, we have:
$d\bs{e}_{a}=-\omb^{b}_{\;\:a}\bs{e}_{b}\;\;\Rightarrow\;\;
\of{D_{\omb}\bs{e}}_{a}=0$.
\Par 
The invariance of definition (\ref{5a}) (whatever the basis one deals
with) with respect to linear transformations,
\be{5b}
\ba{rcl}
\bs{e}\mapsto\ell\bs{e},&\mbox{where}&\ell\in{L}\of{n,\mathbb{R}},\;\:
\mbox{the frame transformation group,}
\ea
\ee
can only be guaranteed if the 1-form $\omega$ transforms as 
\be{6}
\ba{rcl}
\omega&\mapsto&\ell\omega\ell^{-1}+\ell{d}\ell^{-1}.
\ea
\ee
This establishes that $\omega$ indeed serves as a connection for the
frame bundle. Clearly, the linear transformations applied to
an-holonomic frames may form a group as large as $GL\of{n,\mathbb{R}}$. 
But in the approach promoted here, the concept of
metric-compatibility (of the connection) will originate from the covariant
closure of the frames in all bases ({\em regardless\/} of the form of
$\bar{g}$), and the loss of this compatibility will be associated
with a connection deformation process.\Par  
Consider a vector field in its holonomic form,
$\bs{v}=v^{\alpha}\bs{e}_{\alpha}$. This is obviously  
an invariant object. Applying definition (\ref{5}), the ``exterior
derivative'' of $\bs{v}$ reads:
\be{7}
d\bs{v}=\of{dv^{\alpha}-v^{\beta}\omega^{\alpha}_{\;\:\beta}}\bs{e}_{\alpha}
=\of{D_{\omega}v}^{\alpha}\bs{e}_{\alpha}.
\ee
By formulas (\ref{5b})-(\ref{6}), $d\bs{v}$ - the {\em absolute
differential of $\bs{v}$\/} - is a vector $1$-form whose
coefficients are given by $\of{D_{\omega}v}^{\alpha}$. More generally,
the components of the covariant exterior derivative of a tensor
$p$-form $\bs{t}_{p}$ constitute the {\em coordinate tensor\/} of its absolute
differential:
\beq{9}
d\bs{t}_{p}&=&
d\of{{{t_{p}}^{\alpha_{1}\alpha_{2}\ldots}}_{\beta_{1}\beta_{2}\ldots}
{\bs{e}_{\alpha_{1}\alpha_{2}\ldots}}^{\beta_{1}\beta_{2}\ldots}}
\nonumber\\
&\!=\!&
\left(d{{t_{p}}^{\alpha_{1}\alpha_{2}\ldots}}_{\beta_{1}\beta_{2}\ldots}
+\of{-1}^{p+1}
{{t_{p}}^{\gamma\alpha_{2}\ldots}}_{\beta_{1}\beta_{2}\ldots}
\wedge\omega_{\;\:\gamma}^{\alpha_{1}}
+\of{-1}^{p+1}
{{t_{p}}^{\alpha_{1}\gamma\ldots}}_{\beta_{1}\beta_{2}\ldots}
\wedge\omega_{\;\:\gamma}^{\alpha_{2}}+\cdots\right.\nonumber\\
&\!\!&\;\left.
+\of{-1}^{p}{{t_{p}}^{\alpha_{1}\alpha_{2}\ldots}}_{\gamma\beta_{2}\ldots}
\wedge\omega^{\gamma}_{\;\:\beta_{1}}
+\of{-1}^{p}{{t_{p}}^{\alpha_{1}\alpha_{2}\ldots}}_{\beta_{1}\gamma\ldots}
\wedge\omega^{\gamma}_{\;\:\beta_{2}}+\cdots\right)
{\bs{e}_{\alpha_{1}\alpha_{2}\ldots}}^{\beta_{1}\beta_{2}\ldots}
\nonumber\\
&\!=:\!&
{\of{D_{\omega}
{t_{p}}}^{\alpha_{1}\alpha_{2}\ldots}}_{\beta_{1}\beta_{2}\ldots}
{\bs{e}_{\alpha_{1}\alpha_{2}\ldots}}^{\beta_{1}\beta_{2}\ldots}
\;=:\;\of{D_{\omega}\bs{t}}_{p+1}.
\eeq
When people often speak of objects such as vectors or tensors, they
usually 
address, respectively, to coordinate vectors and coordinate tensors
(with respect to a given basis). Here we shall distinguish between
vectors ($\bs{v}$), $p$-forms ($h_{p}$), and tensor $p$-forms
($\bs{t}_{p}$), 
all of which are different types of {\em invariants\/}, and between
coordinate vectors ($v^{\alpha}$), coordinate tensor $p$-forms
(${t_{p}^{\alpha_{1}\alpha_{2}\ldots}}_{\beta_{1}\beta_{2}\ldots}$),
frame elements ($\bs{e}_{\alpha}$), and coframes
($\vartheta^{\alpha}$), all of which transform in a {\em covariant\/}
manner. Note that, whereas $v^{\alpha}$ is {\em scalar-valued\/}, 
$\bs{e}_{\alpha}$ is {\em vector-valued\/}.\Par   
Let us turn back again to the transition functions between bases.
Hitting a transition function
$e_{\alpha}^{a}=\bs{e}_{\alpha}\cdot{\bs{e}}^{a}$ by an exterior
derivative yields  
\be{10}
de_{\alpha}^{a}=-\omega_{\;\:\alpha}^{\beta}e_{\beta}^{a}
+\omb^{a}_{\;\:{b}}e_{\alpha}^{b},
\ee
which can be rewritten as
$\of{D_{\omega\omb}e}_{\alpha}^{a}=0$. In this respect, the
transition functions are regarded as the components of a
covariant-free section in the spliced
bundle $T\Omega^{\otimes\of{1,1}}$, each of whose fiberspaces is a
space product of two tangent spaces, one employs
holonomic basis, the other employs an-holonomic basis,
and there is a distinct connection 1-form for each factor fiber in the
splice. For this reason we prefer to mark the connection in a
non-holonomic basis with a bar, and by thus distinguish it from a
connection in a holonomic basis.\Par 
Multiplying both sides of eq. (\ref{10}) by $\bar{e}^{\gamma}_{a}$
enables to eliminate $\omega$; otherwise, multiplying it by
$\bar{e}^{\alpha}_{c}$ enables to eliminate $\omb$: 
\be{20}\ba{ccc}
\omega^{\gamma}_{\;\:\alpha}=
\bar{e}^{\gamma}_{a}\omb^{a}_{\;\:b}e^{b}_{\alpha}
-\bar{e}^{\gamma}_{a}d{e}_{\alpha}^{a}\,;\;&\;
\omb_{\;\:c}^{a}=
\bar{e}^{\alpha}_{c}\omega_{\;\:\alpha}^{\beta}e_{\beta}^{a}
+\bar{e}^{\alpha}_{c}de_{\alpha}^{a}.
\ea\ee
The connections in the two bases are therefore interrelated through a
$GL\of{n,\mathbb{R}}$ gauge transformation with the transition
functions $\offf{\bar{e}^{\alpha}_{a}}\in{GL}\of{n,\mathbb{R}}$ 
being the group elements. In other words, the connections in the two
bases correspond to two points in two distinct sectors of a {\em single\/} 
gauge orbit in a {\em single\/} general linear group structure.\Par 
From eq. (\ref{10}), the exterior derivative of $g_{\alpha\beta}$,
\beq{30}
dg_{\alpha\beta}\,=\,
d\of{{e}_{\alpha}^{a}{e}_{\beta}^{b}\bar{g}_{ab}}
&\!=\!&-\,
\omega_{\;\:\alpha}^{\gamma}{e}_{\gamma}^{a}{e}_{\beta}^{b}
\bar{g}_{ab}-
\omega_{\;\:\beta}^{\gamma}{e}_{\alpha}^{a}
{e}_{\gamma}^{b}\bar{g}_{ab}\\&&
+\,{e}_{\alpha}^{c}{e}_{\beta}^{b}\omb^{a}_{\;\:{c}}\bar{g}_{ab}
+{e}_{\alpha}^{a}{e}_{\beta}^{c}\omb^{b}_{\;\:{c}}\bar{g}_{ab}
+{e}_{\alpha}^{a}{e}_{\beta}^{b}d\bar{g}_{ab},\nonumber
\eeq
gives rise to the (obvious) relation:
\be{40}
\of{D_{\omega}g}_{\alpha\beta}\;=\;{e}_{\alpha}^{a}{e}_{\beta}^{b}
\of{D_{\omb}\bar{g}}_{ab}.
\ee
However, definition (\ref{5a}) implies that the metric (coordinate)
tensor is a covariant-free object,  
\be{45}
\ba{rcl}
\left.
\ba{l}
dg_{\alpha\beta}=-\omega_{\;\:\alpha}^{\gamma}g_{\gamma\beta}
-\omega_{\;\:\beta}^{\gamma}g_{\gamma\alpha}\\
d\bar{g}_{ab}=-\omb_{\;\:a}^{c}\bar{g}_{cb}-
\omb_{\;\:b}^{c}\bar{g}_{ca}
\ea\right\}
&\Rightarrow&
\of{D_{\omega}g}_{\alpha\beta}\,=\,\of{D_{\omb}\bar{g}}_{ab}\,=\,0.
\ea
\ee
It is precisely by this meaning that the connection was said to be {\em
compatible\/} with the 
metric. Transforming to any other an-holonomic basis won't 
change anything because this would only be a mere gauge transformation. 
This is the property of {\em metricity\/}; the covariant exterior
derivative of the metric tensor vanishes in {\em any\/} basis.\Par 
If $\offf{\bs{e}_{a}}$ form an $x$-independent rectilinear system
everywhere, namely $\bar{g}=\mbox{diag}\of{p,q}=:\eta$ ($p+q=n$),
then, by the lower left equation in (\ref{45}),
$\omb_{ab}=-\omb_{ba}$, and the restriction to orthonormal frames
is established as a symmetry structure with the (pseudo)
rotational group (the isometry group of $\eta$) being the gauge
group. In this case, for any element $o\in{O}\of{p,q}$,   
\be{6b}
\ba{rclcrclcrcl}
\bs{e}&\mapsto&{o}\bs{e},&&
\eta&\mapsto&\of{o\otimes{o}}\eta=\eta,&\mbox{and}&
\omb&\mapsto&{o}\omb{o}^{-1}+o{d}o^{-1}.
\ea
\ee\Par
The connection coefficients in their holonomic version are obtained in 
a standard manner by permuting the indices in
$\partial_{\alpha}g_{\beta\gamma}$ and summing over with alternating 
signs, 
\be{46}
\partial_{\alpha}g_{\beta\gamma}
-\partial_{\beta}g_{\gamma\alpha}
+\partial_{\gamma}g_{\alpha\beta}=
-\omega_{\off{\alpha|\beta|\gamma}}
+\omega_{\off{\beta|\gamma|\alpha}}
-\omega_{\off{\gamma|\alpha|\beta}}
-2\omega_{\gamma\beta\alpha}.
\ee
Using Schouten's convention \cite{S}, $\offf{\alpha\beta\gamma}=
\alpha\beta\gamma-\beta\gamma\alpha+\gamma\alpha\beta$, eq. (\ref{46})
can be put also in the compact form,
\be{47}
\omega_{\gamma\beta\alpha}=-\frac{1}{2}\off{
\partial_{\left\{\alpha\right.}g_{\left.\beta\gamma\right\}}
+\omega_{\offf{\off{\alpha|\beta|\gamma}}}}.
\ee
In a holonomic basis, the anti-symmetric pieces in
$\omega_{\offf{\off{\alpha|\beta|\gamma}}}$ vanish, whence
$\omega_{\gamma\beta\alpha}$ reduces to (minus) the 
Christoffel symbol with the upper index lowered. In its non-holonomic
version formula (\ref{46}) employs $\bar{g}$ and $\omb$, instead of
$g$ and $\omega$. In particular, in a rectilinear system the
$\partial\bar{g}$'s vanish, in which case $\omb$ is purely determined
by the an-holonomy coefficients (see eqs. (\ref{102})-(\ref{103}),
section \ref{s24}).  
%--------------------------
\subsection{\sf The emergence of non-metricity}\label{s22}
%-------------------------
Having seen that metricity is ingrained in definition (\ref{5}), we
realize that in order to avoid it we must modify the definition so as
to make it less restrictive. The simplest sensible modification is to
add a non-homogeneous term in the r.h.s. of the original definition:  
\be{50}
\lim_{\Delta{x}\rightarrow{0}}\off{
\left.\bs{e}_{\alpha}\of{x+\Delta{x}}\right|_{x}-
\left.\bs{e}_{\alpha}\of{x}\right|_{x}}=
-\varpi^{\beta}_{\;\:\alpha}\of{x}
\left.\bs{e}_{\beta}\of{x}\right|_{x}-
\left.\bs{q}_{\alpha}\of{x}\right|_{x},
\ee
where $\bs{q}_{\alpha}\of{x}$ is a vector-valued 1-form (note:
$\varpi\neq\omega$). In our abbreviated notation,
\be{51}
d\bs{e}_{\alpha}=-\varpi_{\;\:\alpha}^{\beta}\bs{e}_{\beta}-\bs{q}_{\alpha},
\ee
which means that the expansion of $d\bs{e}_{\alpha}$ in the basis
$\offf{\bs{e}_{\alpha}}$ should be {\em locally\/} corrected.
As a consequence of this correction, the frames are no longer
covariant-free (with respect to $\varpi$, of course), but instead we
have: 
\be{51a}
\of{D_{\varpi}\bs{e}}_{\alpha}=-\bs{q}_{\alpha}.
\ee 
The $\bs{q}$-terms endow the underlying manifold with 
additional structure (more properly, with a sub-structure) which  
is obviously reflected in the form of the connection coefficients, see
eq. (\ref{60}) ahead. Since each shift $\bs{q}_{\alpha}$ in definition 
(\ref{50}) is a {\em vector-valued\/} object, it can be
expanded in the basis $\offf{\bs{e}_{\alpha}}$:
\be{NM}
\ba{rcl}
\bs{q}_{\alpha}={Q}_{\;\:\alpha}^{\beta}\bs{e}_{\beta}&\Rightarrow&
{Q}_{\alpha\beta}=\bs{e}_{\alpha}\cdot\bs{q}_{\beta},
\ea
\ee
where $Q^{\beta}_{\;\:\alpha}$ is a coordinate tensor 1-form. From
eqs. (\ref{51}) and (\ref{NM}), definition (\ref{5}) is 
back recovered with respect to the composed connection 
$\omega_{\;\:\alpha}^{\beta}= 
\varpi_{\;\:\alpha}^{\beta}+{Q}_{\;\:\alpha}^{\beta}$.\Par  
Eq. (\ref{51}) gives rise to non-metricity through
\be{NM1}
dg_{\alpha\beta}=-\varpi_{\alpha\beta}-\varpi_{\beta\alpha}
-\bs{q}_{\alpha}\cdot\bs{e}_{\beta}-\bs{e}_{\alpha}\cdot\bs{q}_{\beta}
\;\Rightarrow\;
\of{D_{\varpi}g}_{\alpha\beta}=-Q_{\of{\alpha\beta}}
\ee
with $Q_{\of{\alpha\beta}}:=\bs{q}_{\alpha}\cdot\bs{e}_{\beta}+
\bs{e}_{\alpha}\cdot\bs{q}_{\beta}$. Definition (\ref{50}) must hold
also in any non-holonomic basis, 
\be{52}
d\bs{e}_{a}=-\vpb_{\;\:a}^{b}\bs{e}_{b}-\bs{q}_{a}
\ee
and the non-metricity in its an-holonomic version reads:
\be{54}
d\bar{g}_{ab}=-\vpb_{ab}-\vpb_{ba}
-\bs{q}_{a}\cdot\bs{e}_{b}-\bs{e}_{a}\cdot\bs{q}_{b}\;\Rightarrow\;
\of{D_{\vpb}\bar{g}}_{ab}=-Q_{\of{ab}}
\ee
with
$Q_{\of{ab}}:=\bs{q}_{a}\cdot\bs{e}_{b}+\bs{e}_{a}\cdot\bs{q}_{b}$.
Consistency then requires that the $\bs{q}$'s in the holonomic and the
an-holonomic bases should be interrelated via
$\bs{q}_{\alpha}=e_{\alpha}^{a}\bs{q}_{a}$. This requirement, in turn,
implies that the $n$-beins are not necessarily covariant free because 
\be{56}
\of{D_{\varpi\vpb}e}^{a}_{\alpha}=-\bs{q}_{\alpha}\cdot{\bs{e}}^{a}+
\bs{e}_{\alpha}\cdot\bs{q}^{a}=g^{ab}e_{\alpha}^{c}Q_{\off{bc}},
\ee
and $Q_{\off{ab}}$ may not vanish. Nevertheless, the four extra terms
generated in $dg_{\alpha\beta}$ due to (\ref{56}) cancel each other
exactly, therefore eqs. (\ref{30})-(\ref{40}) maintain their original
form, this time, however, with non-vanishing non-metricity:
\be{58}
\of{D_{\varpi}g}_{\alpha\beta}=
e_{\alpha}^{a}e_{\beta}^{b}\of{D_{\vpb}\bar{g}}_{ab}
=-e_{\alpha}^{a}e_{\beta}^{b}Q_{\of{ab}}\equiv-Q_{\of{\alpha\beta}}.
\ee\Par
A spacetime endowed with connection and metric structure
that satisfy eq. (\ref{58}) is called {\em post-Riemannian\/}, and is
frequently denoted by 
$\of{L_{n},g}$. Following the method used in getting eq. (\ref{46}),
the connection coefficients in their holonomic version now satisfy:
\beq{60}
&&\partial_{\alpha}g_{\beta\gamma}
-\partial_{\beta}g_{\gamma\alpha}
+\partial_{\gamma}g_{\alpha\beta}
+Q_{\alpha\of{\beta\gamma}}
-Q_{\beta\of{\gamma\alpha}}
+Q_{\gamma\of{\alpha\beta}}\\
&&\hspace{3cm}=
-\varpi_{\off{\alpha|\beta|\gamma}}
+\varpi_{\off{\beta|\gamma|\alpha}}
-\varpi_{\off{\gamma|\alpha|\beta}}
-2\varpi_{\gamma\beta\alpha}.\nonumber
\eeq
\be{70}
\Rightarrow\;\;\varpi_{\gamma\beta\alpha}\;=\;
-\frac{1}{2}\off{\partial_{\left\{\alpha\right.}g_{\left.\beta\gamma\right\}}
+Q_{\offf{\alpha\of{\beta\gamma}}}+
\varpi_{\offf{\off{\alpha\left|\beta\right|\gamma}}}}.
\ee
This is the well-known Schouten formula, see \cite[page 132]{S}, or
\cite[eq. (3.10.8)]{HMMN}.\footnote{In general,
$\varpi_{\off{\alpha|\beta|\gamma}}\neq0$ even in
a holonomic basis due to the presence of torsion, see
eq. (\ref{122}).} Here, and contrary to the case of
$\omega_{\gamma\beta\alpha}$ - see eq. (\ref{47}) - all the $n^{3}$  
degrees of freedom in the connection are employed, reflecting the
removal of the metricity constraint.\Par
It is important to notice that, due to (\ref{51}), the covariant
exterior derivative of a vector field in an $\of{L_{n},g}$ is
no longer the coordinate vector of its absolute differential,
\be{80}
d\bs{v}\,=\,\of{D_{\varpi}v}^{\alpha}\bs{e}_{\alpha}-
v^{\alpha}\bs{q}_{\alpha}
\ee
(compare this with eq. (\ref{7})); yet, $d\bs{v}$ is still an invariant
because $\bs{q}_{\alpha}$ is covariant. In the more general case, the
absolute differential of a tensor $p$-form (recall eq. (\ref{9}))
involves additional contractions, 
\beq{90}
d\bs{t}_{p}&=&{\of{D_{\varpi}{t_{p}}}^{\alpha_{1}\alpha_{2}\cdots}}_
{\beta_{1}\beta_{2}\cdots}
{\bs{e}_{\alpha_{1}\alpha_{2}\cdots}}^{\beta_{1}\beta_{2}\cdots}
+\of{-1}^{p+1}
{{t_{p}}^{\gamma\alpha_{2}\cdots}}_{\beta_{1}\beta_{2}\cdots}
\wedge{Q}_{\;\:\gamma}^{\alpha_{1}}
{\bs{e}_{\alpha_{1}\alpha_{2}\cdots}}^{\beta_{1}\beta_{2}\cdots}
+\cdots\nonumber\\
&&+\;\of{-1}^{p}{{t_{p}}^{\alpha_{1}\alpha_{2}\cdots}}_{\gamma\beta_{2}
\cdots}\wedge{Q}^{\gamma}_{\;\:\beta_{1}}
{\bs{e}_{\alpha_{1}\alpha_{2}\cdots}}^{\beta_{1}\beta_{2}\cdots}
+\cdots\,.
\eeq
In particular, substituting
$\bs{g}=g_{\alpha\beta}{\bs{e}}^{\of{\alpha\beta}}$ for $\bs{t}_{0}$ 
in (\ref{90}) ($\bs{e}^{\of{\alpha\beta}}$ is a symmetric basis
for $T\Omega^{\otimes2}$, the bundle of rank-2 tensors
over $\Omega$) yields, 
\be{95}
d\bs{g}\,=\,\of{D_{\varpi}g}_{\alpha\beta}{\bs{e}}^{\of{\alpha\beta}}
+g_{\gamma\beta}Q^{\gamma}_{\;\:\alpha}{\bs{e}}^{\of{\alpha\beta}}
+g_{\alpha\gamma}Q^{\gamma}_{\;\:\beta}{\bs{e}}^{\of{\alpha\beta}}
\,=\,0.
\ee
Hence, despite the fact that the {\em metric tensor\/} is not
covariant-free, its absolute differential vanishes just like in the
Riemannian case. In view of this result, and in view of the relation
between the absolute differential of an object, and its covariant
exterior derivative in general, we shall next re-examine 
the concept of parallel displacement in an $\of{L_{n},g}$.
%----------------------------------------
\subsection{\sf A digression on parallel displacements in an
$\of{L_{n},g}$}\label{s23}  
%----------------------------------------
In the following we shall confine ourselves to holonomic bases, but
our arguments, of course, apply to an-holonomic bases as well. Let
$\bs{v}=v^{\alpha}\bs{e}_{\alpha}$ be a vector field whose norm is 
given by $v^{2}=\bs{v}\cdot\bs{v}=v^{\alpha}v^{\beta}g_{\alpha\beta}$.
The exterior derivative of $v^{2}$ reads:
\beq{300}
d\of{\bs{v}\cdot\bs{v}}&=&
\of{D_{\varpi}v}^{\alpha}v^{\beta}g_{\alpha\beta}+
v^{\alpha}\of{D_{\varpi}v}^{\beta}g_{\alpha\beta}-
v^{\alpha}v^{\beta}Q_{\of{\alpha\beta}}\\
&=&D_{\varpi}\of{v^{\alpha}v^{\beta}g_{\alpha\beta}}\,=\,
D_{\varpi}\of{\bs{v}\cdot\bs{v}};\label{302}
\eeq
it displays in a naive manner the breakdown of the Leibnitz rule
with respect to {\em covariant\/} exterior differentiation, because
\be{310}
D_{\varpi}\of{\bs{v}\cdot\bs{v}}\,\neq\,\of{{D}_{\varpi}\bs{v}}\cdot\bs{v}
+\bs{v}\cdot\of{D_{\varpi}\bs{v}}.
\ee
From eq. (\ref{80}), the inequality (\ref{310}) stands in contrast with
$d\of{\bs{v}\cdot\bs{v}}=d\bs{v}\cdot\bs{v}+\bs{v}\cdot{d}\bs{v}$.\Par
The parallel displacement of a vector $\bs{v}$ along a curve with
tangent vector $\bs{u}$ is usually identified with the evaluation of
$D_{\varpi}\bs{v}$ on $\bs{u}$; $\bs{v}_{P}$ is therefore said to
be parallel if it satisfies $D_{\varpi}\bs{v}_{P}=0$. The
inequality (\ref{310}), however, raises the question whether 
parallel displacement is a reliable concept in an $\of{L_{n},g}$.
For, suppose that $\bs{v}_{P}$ has a null norm, namely
$\bs{v}_{P}\cdot\bs{v}_{P}=v_{P}^{\alpha}v_{P}^{\beta}g_{\alpha\beta}=0$.
Then, by eqs. (\ref{300})-(\ref{302}), and due to
$\of{D_{\varpi}v_{P}}^{\alpha}=0$, we have: 
\be{312}
0\,=\,D_{\varpi}\of{\bs{v}_{P}\cdot\bs{v}_{P}}\,=\,
-v_{P}^{\alpha}v_{P}^{\beta}Q_{\of{\alpha\beta}}
\ee
but, apart from one special case (the Weyl-Cartan spacetime),
$Q_{\of{\alpha\beta}}\neq{f}\times{g}_{\alpha\beta}$, where $f$ is a
1-form, leading to an apparent inconsistency.\Par  
Therefore, instead of identifying a parallel field $\bs{v}_{P}$
with a field whose covariant exterior derivative vanishes, it seems
more appropriate here to identify it with a field whose
absolute differential vanishes, namely $d\bs{v}_{P}={0}$. 
$\bs{v}_{P}$ is then said to be {\em absolutely closed\/}. For example, 
metric tensor $\bs{g}$ in eq. (\ref{95}) is absolutely closed, but
not covariantly closed. Of course, in the absence of non-metricity, in
a metric spacetime, the two concepts of covariant closure and absolute
closure become one and the same thing.\Par  
{\em However\/}, in the spirit of eq. (\ref{51a}) we have: 
\be{314}
D_{\varpi}{\bs{v}}:=D_{\varpi}\of{v^{\alpha}\bs{e}_{\alpha}}
=\of{D_{\varpi}v}^{\alpha}\bs{e}_{\alpha}
+v^{\alpha}\of{D_{\varpi}\bs{e}}_{\alpha}
=\of{D_{\varpi}v}^{\alpha}\bs{e}_{\alpha}
-v^{\alpha}\bs{q}_{\alpha}\equiv{d}\bs{v}.
\ee 
It is therefore only in the sense of the {\em definition\/}
$D_{\varpi}\bs{v}:=\of{D_{\varpi}v}^{\alpha}\bs{e}_{\alpha}$ 
that the inequality (\ref{310}) holds true, and that the concepts 
of absolute closure and covariant closure are different from each
other. From the point of view of eq. (\ref{314}) there is no such
difference because we merely decomposed a Riemannian connection into a
post-Riemannian one plus a compensation term; in this respect a
parallel field is unambiguously defined through
$D_{\varpi}{\bs{v}_{P}}=D_{\omega}{\bs{v}_{P}}=d\bs{v}_{P}=0$.  
%--------------------------
\subsection{\sf The inclusion of torsion, and the structural
identities}\label{s24}  
%-------------------------
From this point, and throughout the rest of this work, we will
no longer distinguish between different types of bases. We shall
instead work in a genuine {\em non-holonomic\/} basis (the one attached to a
local observer), and we shall employ {\em Greek\/} 
indices for that basis; we will not mark the connection with a bar, and
we shall use the (bare) letter $g$ for the local metric.\Par 
It is common to think of a coframe element $\vartheta^{\alpha}$ simply as
an object `dual' to the frame element $\bs{e}_{\alpha}$. This
interpretation, however, might mislead:
coframes, as opposed to frames, are by definition Grassmann elements
and are therefore automatically annihilated by $d^{2}$. In order to
sharpen this point, we return for a moment to Riemannian
geometry.\Par 
A second application of $d$ at the frames generates the curvature
2-form,
\be{100}
dd\bs{e}_{\alpha}\,=\,
-\of{d \omega^{\beta}_{\;\:\alpha}+
\omega^{\gamma}_{\;\:\alpha}\wedge\omega^{\beta}_{\;\:\gamma}}
\bs{e}_{\beta}\,=\,-{R}_{\;\:\alpha}^{\beta}\of{\omega}
\bs{e}_{\beta},
\ee
and in general, the $\of{2k}$-th application gives:
\be{100a}
d^{2k}\bs{e}_{\alpha}\,=\,\of{-1}^{k}
R_{\;\:\alpha}^{\beta}\wedge{R}_{\;\:\beta}^{\gamma}\wedge
\cdots\wedge{R}_{\;\:\gamma}^{\delta}\bs{e}_{\delta},
\ee
where $R\equiv{R}\of{\omega}$.\footnote{Note that
$R\of{\omega}\neq{D}_{\omega}\omega\of{=
d\omega+\omega\wedge\omega+\omega\wedge\omega}$; instead,
$R\of{\omega}=d\omega+\omega\wedge\omega=
d\omega+\frac{1}{2}\gb{\omega}{\omega}$.} Therefore, 
$\of{d^{2k}\bs{e}_{\alpha}}\otimes{\bs{e}}^{\alpha}$ is 
the $k$-th order polynomial in the rotational curvature (given up to a
sign), and $\of{d^{2k}\bs{e}_{\alpha}}\cdot{\bs{e}}^{\alpha}$ is
simply its trace. Similarly, if $\bs{v}_{p}$ is a vector $p$-form,
then, by formula (\ref{i3}), and from (\ref{100}):
$d^{2}\bs{v}_{p}=-v_{p}^{\alpha}\wedge{R}_{\;\:\alpha}^{\beta}
\of{\omega}\bs{e}_{\beta}$.\Par
Since a coframe system is required to play a role {\em similar\/} to that
played by a frame system, the exterior derivative of a coframe element
$\vartheta^{\alpha}$ has to be given by 
\be{101}
d\vartheta^{\alpha}=-\vartheta^{\beta}\wedge\omega^{\alpha}_{\;\:\beta}.
\ee
On the other hand, being a Grassmannian element, and because $d$ is a
raising operator in a de-Rham complex, $\vartheta^{\alpha}$
necessarily satisfies the relation 
\be{102}
d\vartheta^{\alpha}=:\frac{1}{2}{C}^{\alpha}_{\;\:\off{\beta\gamma}}
\vartheta^{\beta}\wedge\vartheta^{\gamma},
\ee
where $d\vartheta^{\alpha}=:C^{\alpha}$ is the {\em an-holonomy\/}
2-form.\footnote{$C^{\alpha}$ is not a covariant object; if $\vartheta$
transforms as $\vartheta\mapsto\vartheta\ell^{-1}$, then $C$ transforms as
$C\mapsto{C}\ell^{-1}-\vartheta\wedge{d}\ell^{-1}$.}
Eqs. (\ref{101}) and (\ref{102}) imply the equivalence
\be{103}
\omega_{\off{\beta\left|\alpha\right|\gamma}}
\equiv{C}_{\alpha\off{\beta\gamma}}.
\ee
Now, in Contrast with $d^{2}\bs{e}\neq0$, we have, by definition,
$d^{2}\vartheta=0$. Hence, when $d^{2}$ is applied to coframes, it
generates {\em constraints\/}, rather than new objects as in
eqs. (\ref{100})-(\ref{100a}). For example: if $d$ is applied at
eq. (\ref{101}) it yields the constraint 
\be{110}
0\,=\,dd\vartheta^{\alpha}\,=\,\vartheta^{\beta}\wedge{R}^{\alpha}_{\;\:\beta}
\of{\omega};
\ee
one recognizes formula (\ref{110}) as the $1$-st Bianchi identity in a
Riemannian spacetime $V_{n}$. On the other hand, if $d$ is applied at
eq. (\ref{102}), it gives
\be{111}
dC^{\alpha}_{\;\:\off{\beta\gamma}}
=\frac{1}{2}\of{C^{\alpha}_{\;\:\off{\beta\delta}}
C^{\delta}_{\;\:\off{\gamma\epsilon}}-
C^{\alpha}_{\;\:\off{\gamma\delta}}
C^{\delta}_{\;\:\off{\beta\epsilon}}}\vartheta^{\epsilon},
\ee
which is a variant of the Maurer-Cartan equation associated with the
bundle of orthonormal frames.
However, had we applied $d$ at eq. (\ref{102}), and used
eq. (\ref{101}) (instead of eq. (\ref{102})) for $d\vartheta$, we would
have obtain:
\be{112}
\of{D_{\omega}C^{\alpha}}_{\off{\beta\gamma}}=0,
\ee
which is, in fact, that particular condition on the an-holonomy,
playing a role parallel to the metricity condition on the metric. 
This is the {\em holonomicity condition\/}. Notice that the covariant
differentiation of $C^{\alpha}$ is taken with respect to its {\em 
basespace\/} indices.\Par
We now return to the post-Riemannian case. Owing to the similar role
played by frames and coframes, a consistent extension of the
Riemannian relation (\ref{101})
would be to add an $x$-dependent shift in its r.h.s.:
\be{120}
d\vartheta^{\alpha}=-\vartheta^{\beta}\wedge\varpi_{\;\:\beta}^{\alpha}
-T^{\alpha}. 
\ee
Should definition (\ref{120}) hold in any given basis, the {\em
torsion\/} - $\bs{T}:=T^{\alpha}\bs{e}_{\alpha}$ - must behave as a
vector 2-form. This is 
the place to comment that in a non-holonomic basis, the distinction
between the base sector and the fiber sector becomes rather vague. For 
example, the index $\alpha$ in $d\vartheta^{\alpha}$ in (\ref{120}) is
essentially a fiberspace index although the same index in
$\vartheta^{\alpha}$ is a basespace index; see also eq. (\ref{112}),
and eq. (\ref{123}) below, where a gauge-sector differentiation is
applied to basespace indices.\Par 
Combining eqs. (\ref{102}) and (\ref{120}), the torsion and
the an-holonomy give rise to the following
decomposition of the antisymmetric piece in the connection:
\be{122}
\varpi_{\off{\beta\left|\alpha\right|\gamma}}=
C_{\alpha\off{\beta\gamma}}+
T_{\alpha\off{\beta\gamma}}.
\ee
Furthermore, from eqs. (\ref{102}) and (\ref{120}),
\be{123}
\of{D_{\varpi}C^{\alpha}}_{\off{\beta\gamma}}
\wedge\vartheta^{\beta}\wedge\vartheta^{\gamma}=
C^{\alpha}_{\;\:\off{\beta\gamma}}
T^{\left[\beta\right.}\!\wedge\vartheta^{\left.\gamma\right]},
\ee
which signifies the removal of the constraint on the an-holonomy in a
manner similar to the removal of the metricity constraint on the
metric. This is {\em non-holonomicity\/}, the an-holonomy counterpart of 
non-metricity.\Par 
Let $h=h_{\alpha}\vartheta^{\alpha}$ be a 1-form. Making use of
definition (\ref{120}), the detailed exterior derivative of
$h$ in a spacetime with torsion is given by
\be{124}
dh=\of{D_{\varpi}h}_{\alpha}\wedge\vartheta^{\alpha}-h_{\alpha}T^{\alpha},
\ee
and $d^{2}h=0$ whatsoever.
By construction, $\of{D_{\varpi}h}_{\alpha}$ treats $h_{\alpha}$ in
the same manner it treats it in $\bs{h}=h_{\alpha}\bs{e}^{\alpha}$.
Generally speaking, in the presence of torsion, the 
absolute differential of a scalar-valued $p$-form 
${t}_{p}=t_{\alpha_{1}\cdots\alpha_{p}}
\vartheta^{\alpha_{1}}\wedge\cdots\wedge\vartheta^{\alpha_{p}}$,
%=\frac{1}{p!}t_{\off{\alpha_{1}\cdots\alpha_{p}}}
%\vartheta^{\alpha_{1}}\wedge\cdots\wedge\vartheta^{\alpha_{p}}
taking into account also the derivative of the Grassmann
basis, gets an additional torsional contribution: 
\be{130}
dt_{p}\,=\,\of{D_{\varpi}{t}_{p}}_{\alpha_{1}\cdots\alpha_{p}}   
\vartheta^{\alpha_{1}}\wedge\cdots\wedge\vartheta^{\alpha_{p}}
-p\,{t}_{\alpha_{1}\alpha_{2}\cdots\alpha_{p}}{T}^{\alpha_{1}}
\wedge\vartheta^{\alpha_{2}}\wedge\cdots
\wedge\vartheta^{\alpha_{p}};
\ee
here $\of{D_{\varpi}{t}_{p}}_{\alpha_{1}\cdots\alpha_{p}}   
\vartheta^{\alpha_{1}}\wedge\cdots\wedge\vartheta^{\alpha_{p}}$
treats the (basespace) components of ${t}_{p}$ as if they were the
components of an anti-symmetric tensor field $\bs{t}_{p}=
t_{\alpha_{1}\cdots\alpha_{p}}
\bs{e}^{\off{\alpha_{1}\cdots\alpha_{p}}}$.\Par
As we have already argued, in contradistinction with forms, $d^{2}$
fails to vanish on frames. Instead, in the presence of non-vanishing
non-metricity, 
\be{138}
dd\bs{e}_{\alpha}=
-R_{\;\:\alpha}^{\beta}\of{\varpi}\bs{e}_{\beta}-
\of{D_{\varpi}\bs{q}}_{\alpha}
\ee
(compare this with eq. (\ref{100})). From eq. (\ref{100}), the
(tensorial) equality $-D_{\varpi}\bs{q}=\bs{R}\of{\varpi}$ corresponds
to $\bs{R}\of{\omega}=0.$\Par  
Consider the vector 1-form $\bs{\pi}=\vartheta^{\alpha}\bs{e}_{\alpha}$. 
Its square in a coordinate basis is just the usual line element,
$\pi^{2}=\bs{\pi}\cdot\bs{\pi}=
dx^{\alpha}dx^{\beta}g_{\alpha\beta}$.\footnote{Note that the
line-element, like any cone, is a transvection field \cite{S}, not a
tensorial quantity.} In a 
Riemannian spacetime $d\bs{\pi}$ identically vanishes, and
therefore $d\bs{\pi}^{2}$ absolutely closes; in this sense
$\vartheta^{\alpha}$ and $\bs{e}_{\alpha}$ are bases `dual' to one
another. However, in an $\of{L_{n},g}$, in the presence of torsion and
non-metricity, 
\be{135}
d\bs{\pi}=-T^{\alpha}\bs{e}_{\alpha}+\vartheta^{\alpha}\wedge\bs{q}_{\alpha},
\ee
whence we may conclude that $d\bs{\pi}\neq0$. What seems to be an
exceptional case is when the torsion $\bs{T}$ and the shift
$\bs{q}_{\alpha}$ are mutually interrelated through 
\be{136}
\bs{T}=\vartheta^{\alpha}\wedge\bs{q}_{\alpha}.
\ee
In fact, condition (\ref{136}) is more than an exceptional case. As we
shall explain in section \ref{s31} below, it is a {\em consistency
requirement\/} in a theory in which a 
post-Riemannian spacetime is realized as a deformed Riemannian 
spacetime. Due to eq. (\ref{136}), frames and coframes remain dual
concepts, and the line element $\pi^{2}$ is kept absolutely closed,
also in an $\of{L_{n},g}$.\Par  
There are three basic structural identities in an $\of{L_{n},g}$
- the so-called {\em Bianchi identities\/}. The second of which, and probably
the most familiar one, reads:
\be{140}
dR_{\;\:\alpha}^{\beta}\of{\varpi}=
-{R}_{\;\:\gamma}^{\beta}\of{\varpi}\wedge\varpi_{\;\:\alpha}^{\gamma}
+R_{\;\:\alpha}^{\gamma}\of{\varpi}\wedge\varpi_{\;\:\gamma}^{\beta}
\;\;\Rightarrow\;\;\of{D_{\varpi}R\of{\varpi}}_{\;\:\alpha}^{\beta}=0.
\ee
In a Riemannian spacetime, the same identity for $R\of{\omega}$ 
is equivalent to the absolute closure of 
the curvature tensor, $d\bs{R}=0$. In a post-Riemannian spacetime,
however, the curvature no longer absolutely closes; instead we find
the tensorial equation 
\be{141}
d\bs{R}=\gb{\bs{R}}{\bs{Q}},
\ee
as one may directly infer from formula (\ref{90}).
The other two Bianchi identities, the 1-st and the 3-rd, 
emerge from the nilpotency of $d$ on
$\vartheta^{\alpha}$ and $g_{\alpha\beta}$, respectively:
\beq{142}
0&=&dd\vartheta^{\alpha}\;=\;
\vartheta^{\beta}\wedge{R}_{\;\:\beta}^{\alpha}-\of{D_{\varpi}T}^{\alpha},\\
0&=&ddg_{\alpha\beta}\;=\;
\of{D_{\varpi}Q}_{\of{\alpha\beta}}+R_{\of{\alpha\beta}}.\label{144}
\eeq
In particular, from identity (\ref{142}), the transvected curvature
$R_{\alpha\beta}\wedge\vartheta^{\alpha}\wedge\vartheta^{\beta}$ is
purely of a post-Riemannian origin, and from identity (\ref{144}), so
does the curvature-trace,
$g^{\alpha\beta}R_{\alpha\beta}$.\footnote{See in this respect the 
Appendix B.4 in \cite{HMMN}, where the irreducible pieces of the
curvature in an $\of{L_{n},g}$ are classified; see also the related
discussions in \cite{OVEH,OH}.} \Par 
Finally, being a vector 1-form, $\bs{\pi}$ fails to be annihilated 
by $d^{2}$. Instead we find:
\beq{146}
dd\bs{\pi}&=&\vartheta^{\alpha}\wedge{dd}\bs{e}_{\alpha}\;=\;
-\vartheta^{\beta}\wedge{R}_{\;\:\beta}^{\alpha}\of{\varpi}\bs{e}_{\alpha}
-\vartheta^{\alpha}\wedge\of{D_{\varpi}\bs{q}}_{\alpha}\nonumber\\
&=&-\of{D_{\varpi}T}^{\alpha}\bs{e}_{\alpha}
-\vartheta^{\alpha}\wedge\of{D_{\varpi}\bs{q}}_{\alpha},
\eeq
where we exploited eqs. (\ref{i3}), (\ref{138}), and identity
(\ref{142}). As in eq. (\ref{138}), the
(vectorial 3-form) equality $-D_{\varpi}\bs{T}=
\vartheta\wedge{D}_{\varpi}\bs{q}\,$ corresponds to
$\vartheta\wedge\bs{R}\of{\omega}=0$. 
%----------------------
\section{\sf The post-Riemannian merger of internal symmetries with
spacetime symmetries}\label{s3}  
%----------------------
%----------------------
\subsection{\sf The deformation criterion and the basespace-framespace 
splitting}\label{s31}
%----------------------
The invariant object
$\bs{\pi}=\vartheta^{\alpha}\bs{e}_{\alpha}$ whose square in a
coordinate basis is the infinitesimal line element, is a vector
1-form. The ``tensorial power'' of this object would therefore
be a $p$-form: 
\be{500}
\bs{\pi}_{p}=\vartheta^{\alpha_{1}}\wedge\cdots\wedge
\vartheta^{\alpha_{p}}\bs{e}_{\alpha_{1}}\otimes
\cdots\otimes\bs{e}_{\alpha_{p}}.
\ee  
Now, let us recall
that in order to form the square of $\bs{\pi}$ we employed a dot product
that eventually ended up in a {\em transvection\/}. But this wasn't
an `ordinary' dot product because two 1-forms were mapped to a scalar
(the infinitesimal line element). Let us then formulate this mapping in a
precise language. There exists a map, say {\em odot\/} $\odot$, 
\be{502}
\ba{cc}
\odot:&\mbox{\large$\wedge$}^{p}\of{{T\Omega}^{\otimes{p}}}\times
\mbox{\large$\wedge$}^{p}\of{{T\Omega}^{\otimes{p}}}\;\mapsto\;
\mbox{\large$\wedge$}^{0}\of{\Omega},
\ea
\ee
that takes two rank-$p$ tensor $p$-forms (invariants) into a
transvection (an invariant), and whose realization on a pair of
$\bs{\pi}_{p}$'s is given by  
\be{504}
\ba{rcl}
\bs{\pi}_{p}\odot\bs{\pi}_{p}\:\stackrel{\ms{def}}{=}\:
\vartheta^{\alpha_{1}}\vartheta^{\beta_{1}}g_{\alpha_{1}\beta_{1}}\cdots
\vartheta^{\alpha_{p}}\vartheta^{\beta_{p}}g_{\alpha_{p}\beta_{p}}=
{\pi}_{p}^{2}\,,
&\mbox{and so}&{\pi}_{p}^{2}=\of{\pi^{2}}^{p}.
\ea
\ee
In a coordinate basis, ${\pi}_{p}^{2}$ makes up a ``line-element''
on $T\Omega^{\otimes{p}}$.\Par 
In a Riemannian spacetime metricity guarantees that the line-element
remains invariant against parallel displacement. The post-Riemannian
analogue would therefore be the absolute closure of $\bs{\pi}_{p}$
(any $p\geq1$), and this is identically fulfilled only if
\be{510}
\ba{rcl}
\bs{T}=\vartheta^{\alpha}\wedge\bs{q}_{\alpha},&\mbox{or}&
T^{\alpha}=\vartheta^{\beta}\wedge{Q}_{\;\:\beta}^{\alpha},
\ea
\ee
which is precisely condition (\ref{136}). Eq. (\ref{510}) establishes 
the symmetry structure induced on the frames, with the gauge group
$L\of{n,\mathbb{R}}$ of linear transformations\footnote{Here and in
what follows the linear frame-transformation group
$L\of{n,\mathbb{R}}\subseteq{GL}\of{n,\mathbb{R}}$ is assumed to be
larger than $O\of{n}$.}, on a connection-deformation basis: it is {\em
the\/} necessary requirement that the deformation of the connection,
as was presented in eq. (\ref{120}), where the torsion was introduced,
will be compatible with the deformation of the connection as was
presented in eq. (\ref{51}), where non-metricity was introduced:
\be{520}
\ba{rcl}
d\bs{e}_{\alpha}&=&-\underbrace{\of{\varpi^{\beta}_{\;\:\alpha}+
Q_{\;\:\alpha}^{\beta}}}_{\mbox{$=\omega_{\;\:\alpha}^{\beta}$}}
\bs{e}_{\beta}\;\;\;\Leftrightarrow\;\;\;
\bs{q}_{\alpha}\;=\;Q_{\;\:\alpha}^{\beta}\bs{e}_{\beta}\,,\\
d\vartheta^{\alpha}&=&
-\vartheta^{\beta}\wedge
\underbrace{\of{\varpi_{\;\:\beta}^{\alpha}+Q^{\alpha}_{\;\:\beta}}}
_{\mbox{$=\omega_{\;\:\beta}^{\alpha}$}}\;\;\;\Leftrightarrow\;\;\;
T^{\alpha}\;=\;\vartheta^{\beta}\wedge{Q}_{\;\:\beta}^{\alpha}\,.
\ea
\ee\Par
Therefore, in a picture in which a post-Riemannian spacetime is
regarded as a deformed Riemannian spacetime, eq. (\ref{510}), the
so-called {\em deformation criterion\/},  becomes a constitutive
equation, saying that the torsion furnishes an anti-symmetric
piece for the deformation term. On the other hand, since
$Q_{\of{\alpha\beta}}:=Q^{\gamma}_{\;\;\alpha}g_{\gamma\beta}+
Q^{\gamma}_{\;\:\beta}g_{\gamma\alpha}$, the non-metricity furnishes a
symmetric piece. The coordinate tensor 1-form 
$\varphi^{\beta}_{\;\:\alpha}=-Q^{\beta}_{\;\:\alpha}$ is
known by the name ``{\em distortion 1-form\/}''; it will soon
play a central role in a merging theory, as a host for internal
symmetries.\Par  
The fact that we are dealing with a deformation process has
additional implication that merits few words. Usually, the extension of
the symmetry structure follows from an extension of the gauge group by  
adding 
new generators to the generating Lie algebra (this, of course, may also
require change of representation). The connection 1-form of the
extended structure takes values in the 
extended algebra, and those pieces with values in the {\em added\/}
generators are responsible for the non-linear behavior of the (whole)
connection under a transformation generated in the extension. This is,
however, {\em not\/} an appropriate description when it comes to the 
{\em geometrical\/} formulation of the symmetry structure associated with the
frame bundle. To see this explicitly, note that the non-linear terms
generated in the transformation of $\varpi_{\;\:\alpha}^{\beta}$ under
$L\of{n,\mathbb{R}}$ are coming {\em solely\/} from the
rotational piece, since $\bs{Q}=-D_{\varpi}\bs{g}$ is a
true tensor, and since the derivatives (that generate the non-linear
terms) are found only in the Christoffel symbol. Despite this solid
fact, we will soon show that the {\em merger\/} of spacetime with internal
degrees of freedom {\em does\/} impose non-linear behavior on
the distortion tensor $\bs{Q}$, but it is done in a manner that has
nothing to do with frame transformations.\Par   
The next step that we take, which is crucial to subsequent
developments, is to factor out the basespace part in
$\bs{q}_{\alpha}$, 
\be{530}
\bs{q}_{\alpha}=:-A\otimes\bs{\qq}_{\alpha}=:
-A\otimes\Q_{\;\:\alpha}^{\beta}\bs{e}_{\beta}
\ee
with a 1-form $A$ (${A=A_{\gamma}\vartheta^{\gamma}}$), and a
vector-valued $0$-form $\bs{\qq}_{\alpha}$.\footnote{The 
splitting in (\ref{530}) is natural in the sense that the frame
bundle is {\em locally\/} a tensor product between the basespace and
the frame fibers.} We have deliberately invoked a tensor product in
eq. (\ref{530}), 
for we will soon realize that the 1-form $A$ may possess internal
degrees of freedom {\em without\/} essentially affecting the basic
geometrical structure of spacetime. Hence, from now on, $A$ is
assumed to be a matrix-valued field, $A=A^{i}_{\;j}$. For convenience,
however, we shall often suppress the indices this matrix (or any 
function of it) carries.\Par  
Employing the deformation criterion, eq. (\ref{510}), and as a result
of the splitting in (\ref{530}), the torsion takes the form
\be{540}
\bs{T}\,=\,\of{A\wedge\vartheta^{\alpha}}\otimes\bs{\qq}_{\alpha},
\ee
and the components of the non-metricity tensor read:
\be{542}
\ba{rcl}
Q_{\of{\alpha\beta}}\,=\,-A\otimes\Q_{\of{\alpha\beta}}
&\mbox{with}&
{\Q}_{\of{\alpha\beta}}=\bs{e}_{\alpha}\cdot\bs{\qq}_{\beta}
+\bs{e}_{\beta}\cdot\bs{\qq}_{\alpha}.
\ea
\ee
In what follows, we shall decompose the post-Riemannian connection 
as $\varpi_{\;\:\alpha}^{\beta}:=\omega_{\;\:\alpha}^{\beta}
+\varphi_{\;\:\alpha}^{\beta}$ (instead of decomposing the Riemannian
connection as $\omega_{\;\:\alpha}^{\beta}=
\varpi_{\;\:\alpha}^{\beta}+Q_{\;\:\alpha}^{\beta}$).
From the left equations in 
(\ref{520}), and due to (\ref{530}), the distortion 1-form
splits as: 
\be{546}
\varphi_{\;\:\alpha}^{\beta}=-Q_{\;\:\alpha}^{\beta}
=A\otimes\Q_{\;\:\alpha}^{\beta}.
\ee\Par
Notice that, as opposed to 
$\omega^{\beta}_{\;\:\alpha}\wedge\omega^{\alpha}_{\;\:\beta}\equiv0$, 
the product term
$\varphi^{\beta}_{\;\:\alpha}\wedge\varphi^{\alpha}_{\;\:\beta}$  
vanishes {\em only for an Abelian $A$\/}. Therefore, from now on,
since none of the geometrical objects that depend on $A$ (gradely)
commute with each other, much care is required when handling
calculations; in particular, one must keep an eye on the {\em order\/}
in which terms are arranged in a product. For example: because our
convention has been to put tensor bases always to the right, the
connection in a covariant exterior derivative should always appear to
the {\em right\/} of the object being covariantly differentiated.  
%-----------------
\subsection{\sf Embedding a $\mathbb{C}$-valued non-Abelian internal
symmetry structure within the non-metric sector of an $\of{L_{n},g}$}
\label{s33}  
%-----------------
As we have already implied in section \ref{s31}, we intend to
endow $A$ with internal degrees of freedom. More precisely, we wish
to associate $\varphi\of{A}$ with a connection 1-form in a certain
internal space, equipped with a local symmetry structure acted upon by a
complex-valued gauge group $U\of{N,\mathbb{C}}$, where $N$ being the
dimensionality of $U$.\footnote{Content with little, we could have
taken $U$ to be $\mathbb{R}$-valued, but we can do better than that.} 
Hence we require that under the action of any element $u$ in that
group,\footnote{Focusing here only on the general idea, we shall
suppress all kinds of indices in the meantime.}
\be{570}
\ba{rcl}
\varphi\of{A}&\mapsto&u\varphi\of{A}u^{-1}+udu^{-1}\otimes\bs{1}_{n},
\ea
\ee
where $\bs{1}_{n}$ is the identity element in frame space (recall the
$\bs{\varphi}\of{A}$ is a frame-space tensor). This requirement originates
from the mapping  
\be{572}
\ba{rcl}
\bs{q}\of{A}&\mapsto&u\bs{q}\of{A}u^{-1}-udu^{-1}\otimes\bs{e},
\ea
\ee
see section \ref{s34} for the detailed analysis.\Par
Due to this association, and as a result of the mapping rule
postulated in (\ref{570}), 
the ``encrusted'' post-Riemannian connection\footnote{$\rho\of{U}$ means
that $U$ is in the representation $\rho$; $\bs{1}_{\rho\of{U}}$ is the
unit element in that representation.} 
\be{580}
\varpi\of{\omega,\varphi}=\omega\otimes\bs{1}_{\rho\of{U}}+\varphi\of{A},
\ee 
behaves as an `hyper' connection, gauging independently two local
symmetry structures in a spliced fiber bundle (here one symmetry
refers to the frames, the other refers to the internal space where $A$
lives):  
\be{598}
\ba{rcl}
\forall\,\ell\in{L}:\;\;
\varpi&\mapsto&{\ell}\varpi{\ell}^{-1}
+\ell{d}\ell^{-1}\otimes\bs{1}_{\rho\of{U}}\\
\forall\,u\in{U}:\;\;
\varpi&\mapsto&{u}\varpi{u}^{-1}
+u{d}u^{-1}\otimes\bs{1}_{n}.
\ea
\ee
This type of connection establishes a connectivity on the so-called
{\em foliar bundle\/}, a product bundle whose symmetry structures
interlace in a non-trivial manner. The detailed paradigm of this
merger setup, in its most general form, is given in \cite{M1}.\Par 
Let $O\of{p,q}$ be generated by $\oo\of{p,q}$ ($p+q=n$), let
$L\of{n,\mathbb{R}}$ be generated by $\lf\of{n}$,
and let $U\of{N,\mathbb{C}}$ be generated by
$\uu\of{N}$. The algebraic structure of $\varpi$, as
suggested by formulas (\ref{570})-(\ref{598}), can be put in the
pictorial form: 
\be{604}
\underbrace{\oo\of{p,q}\oplus\overbrace{\off{\lf\of{n}/
\oo\of{p,q}}}^{\mbox{$\otimes\:\uu\of{N}$}}}
_{\mbox{$\lf\of{n}$}}
\ee
meaning that $\varpi$ takes its values in the direct sum,
\be{610}
\varpi\,\in\,\off{\oo\of{p,q}\otimes\bs{1}_{\rho\of{U}}}\oplus
\off{\frac{{\lf\of{n}}}{\oo\of{p,q}}
\otimes\uu\of{N}}.
\ee\Par\noindent
In decomposition (\ref{610}), and in the descriptive scheme (\ref{604}),
$\oo\of{p,q}$ generates pseudo-rotations, and the quotient 
space $\s\of{n}=\lf\of{n}/\oo\of{p,q}$
generates shears and dilations. A detailed analysis of the algebra
$\gl\of{n}\supseteq\lf\of{n}$ is given in \cite{SN1}; here we only
mention that the different sectors in this algebra 
satisfy the inclusion relations 
\be{612}
\ba{ccccc}
\off{\oo,\oo}\subset\oo,&&\off{\s,\s}\subset\oo,&\mbox{and}&
\off{\oo,\s}\subset\s.
\ea
\ee\Par
Substituting $\varpi=\omega+\varphi$ in
$\R\of{\varpi}=d\varpi+\varpi\wedge\varpi$ gives:
\be{620}
\R\of{\varpi}\,=\,R\of{\omega}+{D_{\omega}\varphi}
+\varphi\wedge\varphi\,=\,R\of{\omega}+
\gb{\omega}{\varphi}+R\of{\varphi}
\ee
with $R\of{\varphi}=d\varphi+\varphi\wedge\varphi$; the left
decomposition above manifests covariance with respect to orthonormal
transformations (because $\varphi$ is a frame-space coordinate
tensor), the right one manifests covariance with respect to
internal-space transformations, see eq. (\ref{720}) section
\ref{s34}. Finally, since $R\of{\omega}$ and $\gb{\omega}{\varphi}$
are traceless in frame-space, we have:
\be{630}
\R^{\alpha}_{\;\:\alpha}\of{\varpi}
\equiv{R}^{\alpha}_{\;\:\alpha}\of{\varphi}=
d\varphi^{\alpha}_{\;\:\alpha}+
\varphi^{\beta}_{\;\:\alpha}\wedge\varphi_{\;\:\beta}^{\alpha}.
\ee
This curvature-trace piece is known by the name {\em segmental
curvature\/}.\footnote{\label{f1}I thank Yuri 
Obukhov for notifying me on this matter; in ref. \cite{TT} the word {\em
ric\/} was used instead. This shouldn't be confused with the so-called 
Ricci tensor, $\R_{\beta\gamma}
:=\bs{e}_{\gamma}\rfloor\bs{e}_{\alpha}\rfloor\R^{\alpha}_{\;\:\beta}
=\R^{\alpha}_{\;\:\beta\alpha\gamma}$,
where a tensorial index is contracted with a basespace index. Here
$\rfloor$ designates {\em interior multiplication\/}, see 
\cite[Appendix A.1.3]{HMMN} for a summary of its
properties; $\bs{v}\rfloor{h}$ is also said to be the {\em 
evaluation\/} of the form $h$ over the vector field $\bs{v}$
\cite{Tr}.} We shall come back to 
it later in section \ref{s4} when we'll deal with actions.\Par
In what follows, a post-Riemannian spacetime,
equipped with a metric structure ${g}$, and with the
connection (\ref{580}), will be termed {\em merged spacetime\/}, and
will be denoted by $\of{L_{n}\off{A},g}$.
%----------------------
\subsection{\sf Invariance of the (encrusted) spacetime structural
identities under internal gauge transformations}\label{s34}
%----------------------
According to the prescription presented above, $\bs{q}_{\alpha}$
should possess an internal-space non-linear gauge transformation law 
(recall definition (\ref{530}) and eq. (\ref{572})): 
\be{690}
\ba{rcl}
A\otimes\bs{\qq}_{\alpha}&\mapsto&uAu^{-1}\otimes\bs{\qq}_{\alpha}
+udu^{-1}\otimes\bs{e}_{\alpha}.
\ea
\ee
Multiplying this by $\bs{e}^{\beta}$ gives:
\be{700}
\ba{rcl}
\varphi_{\;\:\alpha}^{\beta}&\mapsto&
u\varphi_{\;\:\alpha}^{\beta}u^{-1}+udu^{-1}\otimes\delta_{\alpha}^{\beta}
\ea
\ee
as required. Notice, however, that the off-diagonal components in
$\varphi^{\beta}_{\;\:\alpha}$ transform {\em covariantly\/} with respect to
internal gauge transformations. In terms of the quantities
$Q^{\beta}_{\;\:\alpha}=-\varphi^{\beta}_{\;\:\alpha}$, and
$Q_{\alpha\beta}=g_{\alpha\gamma}Q^{\gamma}_{\;\:\beta}$,
the mapping in (\ref{700}) reads: 
\be{710}
\ba{rclcrcl}
Q^{\beta}_{\;\:\alpha}&\mapsto&uQ^{\beta}_{\;\:\alpha}u^{-1}-
{u}du^{-1}\otimes\delta_{\alpha}^{\beta}
&\;\Rightarrow\;&
Q_{\alpha\beta}&\mapsto&uQ_{\alpha\beta}u^{-1}-udu^{-1}
\otimes{g}_{\alpha\beta},
\ea\ee
hence the components of the non-metricity tensor transform as
\be{711}
\ba{rcl}
Q_{\of{\alpha\beta}}&\mapsto&uQ_{\of{\alpha\beta}}u^{-1}-2udu^{-1}
\otimes{g}_{\alpha\beta}.
\ea
\ee
In particular, in a {\em Weyl-Cartan merged spacetime\/}
$\of{Y_{n}\off{A},g}$, where we have
$Q^{\beta}_{\;\:\alpha}=-A\otimes\delta^{\beta}_{\alpha}$  
(details are given in section \ref{s42}), the left mapping in
(\ref{710}) translates into   
\be{712}
\ba{rcl}
A&\mapsto&uAu^{-1}+udu^{-1},
\ea
\ee
whence $A$ becomes a typical Yang-Mills (YM) gauge potential.\Par
Under the mapping in (\ref{700}), and since $\omega$ is indifferent to
the internal world where $A$ lives (so that it {\em gradely\/}
commutes with $udu^{-1}$),  
\be{720}
\ba{rcl}
\left.
\ba{rcl}
R\of{\omega}&\mapsto&R\of{\omega}\\
R\of{\varphi}&\mapsto&uR\of{\varphi}u^{-1}\\
\gb{\omega}{\varphi}&\mapsto&u\gb{\omega}{\varphi}u^{-1}
\ea\right\}
&\stackrel{\ms{by (\ref{620})}}{\Longrightarrow}&
\ba{rcl}
\R\of{\varpi}&\mapsto&u\R\of{\varpi}u^{-1};
\ea
\ea
\ee
therefore, $\R^{i}_{\;j}\of{\varpi}$ transforms covariantly also with
respect to internal gauge transformations. For example, the brackets
term in (\ref{720}) transforms as: 
\beq{722}
\gb{\omega}{\varphi}_{\;\:\alpha}^{\beta}
&\mapsto&
u\gb{\omega}{\varphi}_{\;\:\alpha}^{\beta}u^{-1}
+\omega_{\;\:\alpha}^{\gamma}\wedge{u}du^{-1}\otimes
\delta_{\gamma}^{\beta}
+\of{udu^{-1}\otimes\delta_{\alpha}^{\gamma}}
\wedge\omega^{\beta}_{\;\:\gamma}
\;=\;u\gb{\omega}{\varphi}_{\;\:\alpha}^{\beta}u^{-1}.\nonumber\\
\eeq\Par
Two essential questions, however, immediately arise:
\begin{enumerate}
\item
By allowing non-Abelian configurations in the base-part of the
distortion 1-form, aren't we altering the form of the spacetime
structural identities? 
\item
Assuming we don't, are these identities (which have now become
``contaminated'' by internal degrees of freedom) indifferent to internal
gauge transformations? 
\end{enumerate}
We stress that the gauge invariance of the structural identities is an
{\em indispensable\/} consistency 
requirement without which nothing makes sense; had the identities
been sensitive to the gauge, then different gauges would
associate with different structural equations.\Par
In order to give an answer to the first question, we must re-derive the
three identities, this time taking into account the fact that
$\varpi$, $Q$, and $T$ no longer gradely commute with each other (nor
do they commute with themselves).  
A repeated careful computation of $d^{2}\vartheta^{\alpha}$ then
explicitly reveals that the 1-st identity, 
\be{730}
\of{D_{\varpi}T}^{\alpha}=\vartheta^{\beta}
\wedge\R_{\;\:\beta}^{\alpha}\of{\varpi},
\ee 
remains intact while passing from Abelian to non-Abelian
configurations.\Par 
As for the 2-nd Bianchi identity, we still identically have
$D_{\varpi}\R\of{\varpi}=0$ but, as opposed to eq. (\ref{141}), this
time 
\be{732}
d\bs{\R}\,=\,
\of{\R^{\alpha}_{\;\:\gamma}\wedge{Q}^{\gamma}_{\;\:\beta}-
\R^{\gamma}_{\;\:\beta}\wedge{Q}^{\alpha}_{\;\:\gamma}}
{\bs{e}}^{\beta}\otimes\bs{e}_{\alpha}
\,\neq\,\gb{\bs{\R}}{\bs{Q}},
\ee
because $\bs{\R}$ and
$\bs{Q}$ no longer commute (unless $A$ is Abelian).\Par  
The 3-rd structural identity,  
$\of{D_{\varpi}Q}_{\of{\alpha\beta}}=-R_{\of{\alpha\beta}}$
(eq. (\ref{144})), nevertheless, requires more care in deriving it:
one should be careful indeed not to make a slip on the ordering of
non-commuting terms, especially in the expression for the covariant
exterior derivative of $Q_{\of{\alpha\beta}}$. Doing so, we arrive at a
3-rd identity suitable for a merged spacetime: 
\beq{740}
0\;=\;ddg_{\alpha\beta}&\!=\!&
\of{\varpi_{\;\:\gamma}^{\delta}\wedge\varpi_{\;\:\alpha}^{\gamma}
+\varpi_{\;\:\alpha}^{\gamma}\wedge\varpi_{\;\:\gamma}^{\delta}}
g_{\delta\beta}\;+\;
\of{\varpi_{\;\:\gamma}^{\delta}\wedge\varpi_{\;\:\beta}^{\gamma}
+\varpi_{\;\:\beta}^{\gamma}\wedge\varpi_{\;\:\gamma}^{\delta}}
g_{\delta\alpha}
\nonumber\\&&
+\;\of{\varpi_{\;\:\alpha}^{\gamma}\wedge\varpi_{\;\:\beta}^{\delta}
+\varpi_{\;\:\beta}^{\gamma}\wedge\varpi_{\;\:\alpha}^{\delta}}
g_{\gamma\delta}
\,-\,\R_{\of{\alpha\beta}}
\,-\,\of{D_{\varpi}Q}_{\of{\alpha\beta}}
\eeq
or, in a more compact and concise form,
\be{750}
-\of{D_{\varpi}\varphi}_{\of{\alpha\beta}}\,=\,
\gb{\varphi}{\varphi}_{\of{\alpha\beta}}
+\of{\varphi\wedge\varphi}_{\of{\alpha\beta}}
-\R_{\of{\alpha\beta}}.
\ee
For an Abelian connection (distortion), each parentheses in
eq. (\ref{740}) (eq. (\ref{750})) vanish, and we are left with the
old relation, 
eq. (\ref{144}). However, if $\varphi$ is non-Abelian, due to a
non-Abelian $A$, the parentheses terms remain, and eq. (\ref{144})
does get corrected.\Par 
Let us now turn to the acute question of gauge-invariance. Recall
first that $\R\of{\varpi}$ is $U$-covariant (eq. (\ref{720})). In
particular, since $\varpi$ transforms as a $U$-connection
(eq. (\ref{598})), the covariant exterior derivative of the curvature
2-form is also $U$-covariant: 
\be{760}
\ba{rclcrcl}
\of{D_{\varpi}\R}_{\;\,\alpha}^{\beta}
&\mapsto&
u\of{D_{\varpi}\R}_{\;\,\alpha}^{\beta}u^{-1}.
\ea
\ee
Hence, if $D_{\varpi}\R$ vanishes in one gauge (as it identically
does), it will vanish in all gauges; the 2-nd identity is therefore
insensitive to internal gauge transformations.\Par 
From the deformation criterion (eq. (\ref{510})), and from eq.
(\ref{546}), the encrusted torsion (in its coordinate vector form)
reads: 
$T^{\alpha}=\varphi^{\alpha}_{\;\:\beta}\wedge\vartheta^{\beta}$. 
Invoking the internal space transformation rule for
$\varphi^{\alpha}_{\;\:\beta}$ (given in (\ref{700})), reveals
the way $T^{\alpha}$ transforms: 
\be{770}
\ba{rcl}
T^{\alpha}&\mapsto&uT^{\alpha}u^{-1}+udu^{-1}\wedge\vartheta^{\alpha}\,;
\ea
\ee
as a consequence, 
\beq{780}
dT^{\alpha}&\mapsto&du\wedge{T}^{\alpha}u^{-1}+udT^{\alpha}u^{-1}+
uT^{\alpha}\wedge{d}u^{-1}+du\wedge{d}u^{-1}\wedge\vartheta^{\alpha}
-udu^{-1}\wedge{d}\vartheta^{\alpha},\nonumber\\
T^{\beta}\wedge\varpi_{\;\:\beta}^{\alpha}&\mapsto&
uT^{\beta}\wedge\varpi_{\;\:\beta}^{\alpha}u^{-1}+
uT^{\alpha}\wedge{d}u^{-1}-
du\wedge\vartheta^{\beta}\wedge\varpi_{\;\:\beta}^{\alpha}u^{-1}+
du\wedge{d}u^{-1}\wedge\vartheta^{\alpha}.\nonumber\\
\eeq
Substituting now (recall that $d\vartheta^{\alpha}$ gradely commutes with
everything)
\beq{790}
-du\wedge\vartheta^{\beta}\wedge\varpi_{\;\:\beta}^{\alpha}u^{-1}
&=&du\wedge{d}\vartheta^{\alpha}u^{-1}+du\wedge{T}^{\alpha}u^{-1}\nonumber\\
&=&-udu^{-1}\wedge{d}\vartheta^{\alpha}+du\wedge{T}^{\alpha}u^{-1}
\eeq
in the second line of (\ref{780}), and subtracting it from the first
line, cancels all non-linear terms in the transformed
$\of{D_{\varpi}T}^{\alpha}$, leading eventually to a covariant
behavior: 
\be{800}
\ba{rcl}
\of{D_{\varpi}T}^{\alpha}&\mapsto&u\of{D_{\varpi}T}^{\alpha}u^{-1}.
\ea
\ee
Thus, since the curvature as well transforms as an internal-space
tensor (formula (\ref{720})), and since the coframe transforms as an
internal-space scalar, 
the 1-st identity (eq. (\ref{730})) remains intact under internal
gauge transformations.\Par 
We still have to check whether the upgraded 3-rd identity,
eq. (\ref{740}), is gauge-invariant. In fact, it is
sufficient to check it only with respect to its trace in frame space,
and this task is already not too cumbersome. First we write
$g^{\alpha\beta}\of{D_{\varpi}Q}_{\of{\alpha\beta}}=
2dQ^{\alpha}_{\;\:\alpha}+Q^{\of{\alpha\beta}}\wedge{Q}_{\of{\alpha\beta}}$;
then, since the non-Abelian piece in $\varpi_{\alpha\beta}$ is 
$\varphi_{\alpha\beta}=-Q_{\alpha\beta}$, and 
since $Q^{\of{\alpha\beta}}\wedge{Q}_{\of{\alpha\beta}}$
splits into
$2Q^{\alpha\beta}\wedge{Q}_{\alpha\beta}+
2Q^{\alpha\beta}\wedge{Q}_{\beta\alpha}$, 
the trace of eq. (\ref{750}) reduces to the obvious equation,
\be{820}
\R^{\alpha}_{\;\:\alpha}
\,=\,d\varphi^{\alpha}_{\;\:\alpha}+
\varphi^{\beta}_{\;\:\alpha}\wedge\varphi^{\alpha}_{\;\:\beta}
\,:=\,sc\of{\varphi}
\ee
($sc\of{\varphi}$ stands for {\sl segmental curvature\/}, see
eq. (\ref{630})) which manifests (internal-space) gauge invariance by
virtue of the mapping rule for $\varphi$, formula (\ref{700}).\Par
%--------------------
\subsection{\sf Protecting the metric sector against invasions from the
internal world\/}\label{s35} 
%--------------------
The absolute differential of a frame, and the exterior derivative of a
coframe, 
\be{821}
\ba{rcl}
d\bs{e}=-\varpi\bs{e}-\bs{q},&\mbox{and}&
d\vartheta=-\vartheta\wedge\varpi-T,
\ea
\ee
are real-valued because they can always be brought into 
orthogonal form:
$d\bs{e}=-\omega\bs{e}$ and $d\vartheta=-\vartheta\wedge\omega$, 
respectively, and $\omega$ is real-valued by default. The distortion
1-form $\bs{q}\of{A}\cdot\bs{e}^{-1}$ is surly complex-valued but
only through its non-framed entrails, via $A$,
and is therefore of no harm as far as the metric sector is
concerned. Yet, the following claim merits a
conduct: already at the level of the inclusion relations in
(\ref{612}), the elements in the subalgebra $\oo$, and those of the
coset space $\s=\mathfrak{l}/\oo$, mix under commutation
relations. Furthermore, the
hyper algebra $\of{\oo\otimes\bs{1}_{\rho\of{\uu}}}
\oplus\of{\s\otimes\uu}$ is in general not a
Lie algebra since its elements usually do not close under 
brackets. Gauge transformations based on the exponentiation of this
hyper algebra are therefore expected to stain the metric branch
with internal degrees of freedom. One might thus ({\em wrongly\/}) infer
that this already renders the merger concept non-realistic.\Par 
Surely, our theory can {\em not\/}
be based on a hyper symmetry structure whose gauge group $\G$ is
generated by the hyper algebra depicted in (\ref{604}). Instead we
base it on the {\em merging\/} of the underlying two symmetries, put
in the form of a spacetime model. However, this can be established
only if the transformations at the frame sector and those of the
internal world remain independent of each other whatsoever (namely
commute). \Par  
Let $\theta\of{x}$ be a local angle with values in the hyper algebra
(\ref{604}), 
\be{830}
\theta
\,\in\,\off{\oo\of{p,q}\otimes\bs{1}_{\rho\of{U}}}\oplus
\off{\frac{{\lf\of{n}}}{\oo\of{p,q}}
\otimes\uu\of{N}};
\ee
let also the {\em Capital trace\/} $\Tr{\ast}$ designate a trace
taken in the representation space of $U$, and the {\em lowercase
trace\/} $\tr{\ast}$ designate a trace taken in frame space. 
Then, {\em assuming\/} that $\lf$ and $\uu$ each carrying a trace, we
assign: 
\beq{840}
\ell\of{\theta}\;:=\;\exp{\mbox{Tr}\,{i\theta}}
\;\in\;{L}\of{n,\mathbb{R}};&&
u\of{\theta}\,:=\,\exp{\mbox{tr}\,{i\theta}}
\;\in\;{U}\of{N,\mathbb{C}}.
\eeq
The assignments made here guarantee that frame
transformations will never involve internal degrees of freedom, and
that internal transformations will never interfere with the
frames; furthermore, the two transformations in (\ref{840}) are 
independent of each other. In this way the segregation
between the metric and the non-metric sectors is preserved under any
of these transformations despite the fact that in both cases the
transformation is compensated by a {\em single common\/} connection
$\varpi$, with values in the hyper algebra (\ref{604}).\Par  
The true symmetry of the merger has thus been set to be
$L\of{n}\times{U}\of{N}$, 
rather than the grand group $\G$, whose elements are given by
$e^{i\theta}$. According to assignments (\ref{840}), the presence of
trace pieces in $\uu$ enables shear and dilation transformations
at the frames sector, whereas the dilation elements in $\lf$ give
rise to internal gauge transformations. %\footnote{The dilations are
%the $n$ non-zero trace pieces in $\gl\of{n}$, corresponding 
%to stretching (or contraction) along the $n$ directions in
%spacetime. This is most easily seen in the standard basis where
%the generators of $\gl\of{n}$ are realized by the $n^{2}$ matrices
%$(M_{\alpha}^{\beta})_{\gamma}^{\delta}=  
%i\delta_{\alpha}^{\delta}\delta_{\gamma}^{\beta}$; the non-zero trace
%elements are those $n$ diagonal matrices in which one entry in the
%diagonal is filled. The center in this algebra, however, contains only
%one element; it is the volume changing generator, given by the direct
%sum $\bigoplus_{\alpha}(M_{\alpha}^{\alpha})_{\gamma}^{\delta}= 
%\delta_{\gamma}^{\delta}$.} 
Otherwise, in the absence of trace pieces in $\uu$, the left-hand
exponent in (\ref{840}) reduces to local (pseudo-) rotations, while in
the absence of dilation elements in $\lf$ (and since $\oo$ is already
traceless in framespace) the right-hand exponent in (\ref{840})
reduces to the identity element.  
%----------------------
\section{\sf Merger of Yang-Mills interactions with gravity}\label{s4}  
%----------------------
%---------------------
\subsection{\sf Selecting an appropriate action\label{s41}}
%---------------------
Our aim now is to select a suitable action for the dynamics of the
merged spacetime, one that economically encompasses the two interlaced
symmetries. There is probably no unique prescription for that task,
but we surly must follow two basic requirements:     
\begin{enumerate}
\item
The action must be invariant under frame transformations;
\item
It must also be invariant under internal gauge transformations.
\end{enumerate}
\Par
With these guidelines in mind, and knowing in advance the outcome of
our choice, we proceed along the lines of \cite{TT} and choose
a rather minimal 
option:\footnote{This type of action, and some variants alike, have 
also been introduced in \cite{HM}.} 
\beq{1002}
S_{LC}\of{g,\omega,\varphi}&=&\int_{\Omega}\mbox{Tr}\off{\R_{\alpha\beta}
\wedge\star\of{\frac{1}{\ell^{n-2}}\vartheta^{\alpha}\wedge\vartheta^{\beta}+
\frac{1}{\ell^{n-4}}g^{\alpha\beta}\R_{\;\:\gamma}^{\gamma}}};
\eeq
here $\star$ stands for the Hodge-star map, the parameter $\ell$,
having the dimension of length, has been installed with the prescribed 
powers, in order to make
the two term in the integrand dimensionless\footnote{This is, of
course, the origin of the gravitational constant in $n=4$.}, and the
capital trace is taken in the representation space of $U$.   
The $\star\of{\ast}$ term in (\ref{1002}) should be
contemplated as a generalized {\em excitation tensor\/} \cite{HMMN};
being adjusted to post-Riemannian geometry, it now contains a {\em
symmetric\/} fiber piece,
$g^{\alpha\beta}=\bs{e}^{\alpha}\cdot\bs{e}^{\beta}$, in addition 
to the traditional anti-symmetric one,
$\vartheta^{\alpha}\wedge\vartheta^{\beta}$.  \Par 
The action functional (\ref{1002}) decomposes into two terms,   
\beq{1010}
S_{LC}\of{g,\omega,\varphi}&=&\frac{1}{\ell^{n-2}}\int_{\Omega}
\Trr{\R_{\alpha\beta}\of{\omega,\varphi}\wedge\star
\of{\vartheta^{\alpha}\wedge\vartheta^{\beta}}}\;+\;
\frac{1}{\ell^{n-4}}\int_{\Omega}\Trr{sc\of{\varphi}
\wedge\star\,{sc}\of{\varphi}},\nonumber\\
\eeq
the first of which {\em resembles\/} the gravitational
Einstein-Hilbert (EH) action in the absence of matter sources, and
will thus be identified with the gravitational  
branch of the merger. The second integral, on the other hand, 
will be associated with internal space dynamics; in one particular
case it will eventually generate a YM source for
energy-momentum in Einstein's gravitational theory.\Par 
Clearly, we could have added an additional $\R^{\alpha\beta}$ 
term into the excitation tensor in
(\ref{1002}). A curvature square term in the action would have
represent contributions that dominate at high curvature,\footnote{The
absence of this term therefore characterizes the low-curvature (or
classical) limit of the theory, hence the subscript ``$LC$'' attached
to $S_{LC}\of{\omega,\varphi}$ in (\ref{1002})-(\ref{1010}).}  
and may possibly become relevant deep in the quantum regime. 
Apart from the Weyl-Cartan case, it will produce strong coupling 
between $\omega$-dependent and $\varphi$-dependent terms; 
in any situation, it will also give rise to the presence of the norm,
\be{1011}
\int_{\Omega}{R^{\alpha}_{\;\:\beta}\of{\omega}
\wedge\star{R}^{\beta}_{\;\:\alpha}\of{\omega}}
\ee
namely, to the existence of a YM-type term for the gravitational
field, in addition to the EH term. Such high-curvature corrections
(even if they can be justified on theoretical grounds) will not be
discussed in the sequel. 
%--------------------------
\subsection{\sf The Weyl-Cartan merger: structure, features, and
dynamics}\label{s42} 
%------------------------- 
The Weyl-Cartan spacetime is the simplest extension of a Riemannian
spacetime that goes beyond the constraints of metricity and
holonomicity. It possesses post-Riemannian geometry with non-vanishing
torsion and non-metricity generated by a minimal number of non-metric
degrees of freedom. Our objective in this section is to show explicitly
that the Weyl-Cartan {\em merged\/} spacetime $\of{Y_{n}\off{A},g}$
provides a comprehensive (classical) framework for the description of
the (bosonic sector of the) present days low energy physics.\Par 
In the Weyl-Cartan spacetime, the vector-valued $0$-form
$\bs{\qq}_{\alpha}$ (defined in eq. (\ref{530})) is taken to be equal to
the frame field $\bs{e}_{\alpha}$ itself,
\beq{1021}
\bs{\qq}_{\alpha}\equiv\bs{e}_{\alpha}\,;
&\mbox{hence, from eqs. (\ref{530}) and (\ref{546})}:&
\;\,
\Q_{\;\:\alpha}^{\beta}=\,\delta_{\alpha}^{\beta}\;\;\Rightarrow\;\;
\varphi^{\beta}_{\;\:\alpha}=A\otimes\delta^{\beta}_{\alpha}.
\nonumber\\
\eeq
It then follows that the coset space
$\s\of{=\lf/\oo}$  
in association (\ref{610})
contains only the identity element in frame space; $\varpi$ takes
its values $\in\off{\oo\otimes\bs{1}_{\rho\of{U}}}\oplus
\off{\bs{1}_{\rho\of{O}}\otimes\uu}$, the Lie algebra of the direct
product group $O\times{U}$, and therefore has the algebraic
structure of a connection in a Whitney product of two vector
bundles \cite{NS}.\Par
The assignment in (\ref{1021}) leads to particularly simple
expressions for the torsion and non-metricity
(eqs. (\ref{540}) and (\ref{542}), respectively):
\beq{1022}
\ba{rcl}
T^{\alpha}\;=\;A\wedge\vartheta^{\alpha},&&
Q_{\of{\alpha\beta}}\;=\;-2A\otimes{g}_{\alpha\beta}\,;
\ea
\eeq
it then follows that transvection terms of the form:  
$\vartheta_{\alpha}\wedge{T}^{\alpha}$, $T_{\alpha}\wedge{T}^{\alpha}$, 
and $Q_{\of{\alpha\beta}}\wedge{T}^{\alpha}\wedge\vartheta^{\beta}$
(which we discuss in the Appendix), identically vanish.\Par 
The segmental curvature, eq. (\ref{630}), now takes the form of 
a YM curvature, 
\be{1026}
sc\of{\varphi}\,=\,n\of{dA+A\wedge{A}}\,=\,:nF\of{A},
\ee
in which case the integrand in the second term of
$S_{LC}\of{g,\omega,\varphi}$ in (\ref{1010}) makes up a free YM
Lagrangian for the potential $A$,
\be{1028}
\Tr{sc\wedge\star{sc}}=n^{2}\,\Tr{\,F\wedge\star{F}}.
\ee
By formula (\ref{710}), and as has already been stated in (\ref{712}),
$A$ and $F$ satisfy the YM gauge transformation
laws,
\be{1040}
\ba{rclcrcl}
A&\mapsto&uAu^{-1}+udu^{-1},&&F&\mapsto&uFu^{-1}.
\ea
\ee
Since the internal branch of the merger acquires the exact content of a
classical YM gauge theory, we may now explicitly assign:
\be{1050}
\ba{rcl}
A^{i}_{\;j}\,:=\,A_{\alpha}^{a}\of{I_{a}}^{i}_{\;j}\vartheta^{\alpha}
&\Rightarrow& 
F^{i}_{\;j}\,=\,\frac{1}{2}F^{a}_{\off{\alpha\beta}}\of{I_{a}}^{i}_{\;j}
\vartheta^{\alpha}\wedge\vartheta^{\beta},
\ea
\ee 
where the $I_{a}$'s stand for the generators of
$U$.\footnote{Hence, if $\U_{p}$ is a frame-space scalar
$p$-form, transforming as the components of a tensor under the action
of $U$, namely $\forall\;u\in{U}:\;\U_{p}\;\mapsto{u}\U_{p}u^{-1}$, 
then, according to (\ref{1040}), the covariant
exterior derivative of $\U_{p}$ with respect to the connection $A$ 
is given by 
\[
\ba{rcl}
\of{D_{A}\U}_{p+1}&\!\!:=\!\!&d\U_{p}+A\wedge\U_{p}+\of{-1}^{p+1}\U_{p}
\wedge{A}=d\U_{p}+\gb{A}{\U_{p}},\\
\mbox{and}\;\;
\of{D_{A}D_{A}\U}_{p+2}&\!\!=\!\!&\off{F\of{A},\U_{p}}.
\ea
\]\label{f2}}
\Par
From $\Q_{\;\:\alpha}^{\beta}=\delta_{\alpha}^{\beta}$, and
because $A$ gradely commutes with $\omega$, the brackets 
$\gb{\omega}{\varphi}$ in (\ref{620}) identically vanish, and the
curvature of the Weyl-Cartan merger decomposes as:
\be{1060}
\R_{\;\:\alpha}^{\beta}\of{\varpi}\,=\,
R_{\;\:\alpha}^{\beta}\of{\omega}\otimes\bs{1}_{\rho\of{U}}
+F\of{A}\otimes\delta_{\alpha}^{\beta},
\ee
or in its ``lowercase'' form: ${\R_{\alpha\beta}\of{\varpi}=
R_{\alpha\beta}\of{\omega}\otimes\bs{1}_{\rho\of{U}}
+F\of{A}}\otimes{g}_{\alpha\beta}$.\footnote{Clearly, since
$\varpi\in\off{\oo\otimes\bs{1}_{\rho\of{U}}}\oplus 
\off{\bs{1}_{\rho\of{O}}\otimes\uu}$, it follows that
$\R\of{\varpi}\in\off{\oo\otimes\bs{1}_{\rho\of{U}}}\oplus
\off{\bs{1}_{\rho\of{O}}\otimes\uu}$ as well.}
Decomposition (\ref{1060}) can also be viewed as if the Riemannian
curvature went through - what we call - {\em projective non-Abelian deformation\/}:
from an internal space perspective, since $F\of{A}$
identically satisfies the Bianchi identity with respect to $A$ (see
footnote \ref{f2}),
\be{1070}
D_{A}F\of{A}\,=\,dF\of{A}+A\wedge{F}\of{A}-F\of{A}\wedge{A}\,=\,0,
\ee
the deformation term in decomposition (\ref{1060}) closes with
respect to $D_{A}$. \Par 
For an Abelian gauge field, the deformation term in eq. (\ref{1060}),
$F\of{A}\otimes\bs{1}_{n}=dA\otimes\bs{1}_{n}$,
becomes a proper projective element \cite[eq. (3.11.8)]{HMMN}.
In particular, in four dimensions, and after renaming the gauge field,
$A\rightarrow-\frac{1}{4}A$, one encounters the triplet    
\be{1082}
\bs{Q}\;\,=\;\,\frac{1}{2}A\otimes\bs{g},
\;\;\;\;\;
\bs{T}\;\,=\;\,
-\frac{1}{4}\of{A\wedge\vartheta^{\alpha}}\otimes\bs{e}_{\alpha},
\;\;\;\;\;
sc\of{A}\;\,=\;\,-dA,
\ee
which is nothing but the Teyssandier-Tucker vacuum configuration, 
discovered in '96 \cite{TT}; in their seminal paper Teyssandier and
Tucker proposed the action (\ref{1010}) (in four dimensions, with an
Abelian gauge field $A$, and equipped 
with a coupling constant also in front of the second term) and looked
for a configuration that {\em extremes\/} it.\Par 
In addition to identity (\ref{1070}), and due to
$D_{\varpi}\R\of{\varpi}=0$, we have one more curvature identity,
namely the 2-nd Bianchi identity associated with the rotational
subgroup, 
\be{1072}
D_{\omega}R\of{\omega}=0,
\ee
and essentially {\em two\/} more torsional identities:
\beq{1074}
\of{D_{\varpi}T}^{\alpha}&=&d\of{A\wedge\vartheta^{\alpha}}-
\of{A\wedge\vartheta^{\beta}}\wedge\varpi^{\alpha}_{\;\:\beta}
%\nonumber\\&=&
\;=\;dA\wedge\vartheta^{\alpha}-A\wedge\of{d\vartheta^{\alpha}+
\vartheta^{\beta}\wedge\varpi_{\;\:\beta}^{\alpha}}\nonumber\\
&=&dA\wedge\vartheta^{\alpha}+A\wedge{T}^{\alpha}
%\nonumber\\&=&
\;=\;F\of{A}\wedge\vartheta^{\alpha},
\eeq
whence, due to eqs. (\ref{730}) and (\ref{1060}),
\be{1073}
\vartheta^{\beta}\wedge{R}_{\;\:\beta}^{\alpha}\of{\omega}\;=\;0.
\ee\Par
From a geometric (and aesthetic) viewpoint, the interrelations
between the internal  
gauge field $A$ and the geometry of the $\of{Y_{n}\off{A},g}$
spacetime may concisely be summarized in a triplet of pairs of
equivalent objects: 
\be{1076}
\ba{rclcrclcrcl}
-A\otimes\bs{g}&\leftrightarrow&\bs{Q},&&
A\wedge\bs{\pi}&\leftrightarrow&\bs{T},&\mbox{and}&
{F}\of{A}\wedge\bs{\pi}&\leftrightarrow&
D_{\varpi}\bs{T}\,
\ea
\ee
(recall that $\bs{\pi}:=\vartheta^{\alpha}\bs{e}_{\alpha}$);
the YM gauge field can therefore be interpreted as a non-Abelian
source for torsion and non-metricity in a $\of{Y_{n},g}$ 
spacetime.\footnote{In ref. \cite{GH} (bottom of page 9) we found a
stinging remark saying, inter alia, that the torsion cannot be closely
related to electromagnetism, nor to any other non-gravitational
(gauge) field. This, however, seems to be true only as long as one
considers the {\em vectorial\/} component of the torsion. Nothing 
negates close relations between the basespace components of the
torsion and the gauge fields of an internal symmetry.}\Par
Let us next compute $\of{D_{\varpi}Q}_{\of{\alpha\beta}}$. Since 
$Q_{\of{\alpha\beta}}=-2A\otimes{g}_{\alpha\beta}$, and from the
relation $dg_{\alpha\beta}=-\varpi_{\alpha\beta}-\varpi_{\beta\alpha}
-Q_{\of{\alpha\beta}}$ we find:
\beq{1078}
\of{D_{\varpi}Q}_{\of{\alpha\beta}}&=&
-2\of{dA\otimes{g}_{\alpha\beta}-A\wedge{d}g_{\alpha\beta}-
A\wedge\varpi_{\alpha\beta}-A\wedge\varpi_{\beta\alpha}}\nonumber\\
&=&-2\of{dA\otimes{g}_{\alpha\beta}+A\wedge{Q}_{\of{\alpha\beta}}}
\;=\;-2F\of{A}\otimes{g}_{\alpha\beta}+6A\wedge{A}\otimes{g}_{\alpha\beta},
\nonumber\\
\eeq
which is seen to hold in agreement with the 3-rd identity,
eq. (\ref{740}). Transvecting this with $g^{\alpha\beta}$ (namely,
taking the trace in frame-space) gives,
\be{1080}
\ba{rcl}
g^{\alpha\beta}\of{D_{\varpi}Q}_{\of{\alpha\beta}}=
-2n\off{F\of{A}-3A\wedge{A}};&\mbox{also,}&
g^{\alpha\beta}Q_{\of{\alpha\beta}}=-2nA.
\ea
\ee
These two may now be substituted into the formula for the
($\of{L_{n}\off{A},g}$-upgraded) 
{\em non-Abelian Stokes' theorem\/} for the non-metricity-trace,
\be{1088}
\int_{\Sigma}g^{\alpha\beta}\of{D_{\varpi}Q}_{\of{\alpha\beta}}
-\int_{\Sigma}
Q^{\of{\alpha\beta}}\wedge{Q}_{\of{\alpha\beta}}
\,=\,\oint_{\partial\Sigma}g^{\alpha\beta}Q_{\of{\alpha\beta}},
\ee
which we develop in Appendix \ref{a12} (here $\Sigma$ stands for a
2-domain that can be covered by a {\em single\/} patch),
to yield the (obvious) result, 
\be{1092}
\int_{\Sigma}dA=\oint_{\partial\Sigma}A;
\ee
one may consider this result as a nice consistency checking.\Par 
Since $\R_{\alpha\beta}\of{\varphi,\omega}=R_{\alpha\beta}\of{\omega}
\otimes\bs{1}_{\rho\of{U}}+F\of{A}\otimes{g}_{\alpha\beta}$, and due
to ${g}_{\alpha\beta}
\star\of{\vartheta^{\alpha}\wedge\vartheta^{\beta}}\equiv{0}$,
the gravitational term in the r.h.s. of (\ref{1010}) reduces to the 
EH action of General Relativity (GR); on the other hand, 
the internal term reduces to free YM (eq. (\ref{1026})). Hence, our
proposed $\of{Y_{n}\off{A},g}$ action takes the compelling form
\beq{1090}
S_{LC}\of{g,\omega,A}
&=&\frac{\dim\rho}{\ell^{n-2}}\int_{\Omega} 
R_{\alpha\beta}\of{\omega}\wedge\star
\of{\vartheta^{\alpha}\wedge\vartheta^{\beta}}\;+\;
\frac{n^{2}}{\ell^{n-4}}\int_{\Omega}\Trr{F\of{A}\wedge\star{F}\of{A}}
\,,\nonumber\\ 
\eeq
with
$R\of{\omega}=d\omega+\omega\wedge\omega$, and $F\of{A}=dA+A\wedge{A}=
\of{D_{\omega}A}_{\alpha}
\wedge\vartheta^{\alpha}+A\wedge{A}$.\footnote{Clearly, in
a holonomic basis $\of{D_{\omega}A}_{\alpha}\wedge\vartheta^{\alpha}$
reduces to $dA_{\alpha}\wedge\vartheta^{\alpha}$ due to
$d\vartheta^{\alpha}=0$.} Needless to elaborate, the action
(\ref{1090}) describes a (non-Abelian) YM gauge field in the presence
of Einstein's gravity in $n$ dimensions.\Par  
The equations of motion that come out by extremising
$S_{LC}\of{g,\omega,A}=S_{LC}\of{\vartheta,\omega,A}$ with respect to
$g_{\alpha\beta}$ are known 
to treat the YM curvature as an internal-space source for energy
momentum in Einstein's equation; this is a standard procedure.
Alternatively, we may vary the action in (\ref{1090}) with
respect to the {\em coframes\/} to get the very same
result.\footnote{I thank Nico Giulini for explaining me this method of 
obtaining the Einstein equation.} The latter method, however, reveals
interesting dynamical features in a transparent manner, hence we 
shall now work it out in details. Since $R\of{\omega}$ and $F\of{A}$
{\em do not depend on the choice of a coframe\/}, what we only need is
a prescription for the variation of forms mapped under the
Hodge star. Such prescription has recently been developed by Muench,
Gronwald, and Hehl \cite[pages 8-9]{MGH}: let $\phi$ be an arbitrary
$p$-form (not carrying values in a Lie algebra). Then, 
\beq{1091}
\delta\star\phi&=&\star\;\delta\phi\,+\,\delta\vartheta^{\alpha}\wedge
\of{\bs{e}_{\alpha}\rfloor\star\phi}\,-\,
\star\off{\delta\vartheta^{\alpha}\wedge\of{\bs{e}_{\alpha}\rfloor\phi}}
\nonumber\\&&
+\;\delta{g}_{\alpha\beta}\off{\vartheta^{\left(\alpha\right.}\wedge\of{
\bs{e}^{\left.\beta\right)}\rfloor\star\phi}
\,-\,\frac{1}{2}g^{\alpha\beta}\star\phi},
\eeq
where $\rfloor$ denotes interior multiplication (see footnote
(\ref{f1})).\Par 
We shall now vary the {\em coframes\/} in (\ref{1090}), leaving all
other fields in the action - including the components of the metric
tensor - untouched (so that $\delta{g}_{\alpha\beta}=0$). Using
$\Tr{F\wedge\star{F}}=\sum_{a}F^{a}\wedge\star{F}^{a}$ with
$a=1\cdots\mbox{dim}\,U$, the evaluation formula 
$\bs{e}_{\alpha}\rfloor\vartheta^{\beta}= 
\delta_{\alpha}^{\beta}\,\Rightarrow\,
\bs{e}^{\alpha}\rfloor\vartheta^{\beta}=g^{\alpha\beta}$,
and the identity
$\int_{\Omega}\sigma\wedge\star\,\phi=\int_{\Omega}\phi\wedge\star\,\sigma$, 
which holds for any two scalar-valued forms $\sigma$ and $\phi$ of
equal degree, we obtain:  
\beq{1093}
\delta_{\vartheta}\int_{\Omega}R_{\alpha\beta}
\wedge\star\of{\vartheta^{\alpha}\wedge\vartheta^{\beta}}
&=&\int_{\Omega}\delta\vartheta^{\gamma}\wedge{R}_{\alpha\beta}\wedge
\bs{e}_{\gamma}\rfloor\star\of{\vartheta^{\alpha}\wedge\vartheta^{\beta}},\\
\delta_{\vartheta}\int_{\Omega}\Tr{F\wedge\star{F}}&=&\int_{\Omega}
\delta\vartheta^{\gamma}\wedge\mbox{Tr}\off{F\wedge\of{\bs{e}_{\gamma}\rfloor
\star{F}}-\of{\bs{e}_{\gamma}\rfloor{F}}\wedge\star\,{F}}.\label{1093a}
\eeq
Hence, as the variation $\delta\vartheta^{\gamma}$ is arbitrary, 
and since $\bs{e}_{\alpha}\rfloor\star\sigma\;=\;
\star\of{\sigma\wedge\vartheta_{\alpha}}$, we finally
arrive at the required equation(s) of motion:
\be{1094}
R_{\alpha\beta}\wedge\star
\of{\vartheta^{\alpha}\wedge\vartheta^{\beta}\wedge\vartheta^{\gamma}}\,=\,
-\kappa\,\mbox{Tr}\off{F\wedge\star\of{F\wedge\vartheta^{\gamma}}-
\of{\bs{e}^{\gamma}\rfloor{F}}\wedge\star\,{F}},
\ee 
with the constant
$\kappa=n^{2}\ell^{2}/\mbox{dim}\rho$ having the dimension of
length-square. The l.h.s. of (\ref{1094})
is the {\em Einstein $\of{n-1}$-form\/} (see \cite{H} for more details
on this object); the r.h.s. contains the {\em  
energy-momentum current\/} associated with the YM gauge field,
\be{1095}
\Sigma^{\gamma}:=
\mbox{Tr}\off{F\wedge\star\of{F\wedge\vartheta^{\gamma}}-
\of{\bs{e}^{\gamma}\rfloor{F}}\wedge\star\,{F}}.
\ee 
As $\vartheta^{\alpha}\wedge\of{\bs{e}_{\alpha}\rfloor\sigma}=p\sigma$
for any $p$-form $\sigma$, the YM energy-momentum current satisfies:
\be{1096}
\vartheta^{\gamma}\wedge\Sigma_{\gamma}=
\of{n-4}\Tr{F\wedge\star{F}},
\ee
and it vanishes {\em only at $n=4$\/}. The vanishing of 
$\vartheta^{\gamma}\wedge\Sigma_{\gamma}$ in four dimensions
corresponds to the tracelessness of the standard energy momentum
tensor in Einstein's gravity - see in this respect the Abelian
treatment of ref. \cite{HO}.\Par  
Taking the variation of $\LL_{LC}\of{\vartheta,\omega,A}$ with respect
to the YM gauge field is a standard procedure. The result (in the
absence of spinorial matter or any external currents) is given by the
so-called YM {\em vacuum\/} equation, 
\be{1099}
D_{A}\star{F}\,=\,0.
\ee
In four dimensions the curvature eigenforms of the Hodge star map,
$\star{F}=\alpha{F}$, where $\alpha$ is any scalar, are obvious
solutions of eq. (\ref{1099}) due to the Bianchi identity for $F$,
$D_{A}F={0}$. It is interesting to note that these special solutions
correspond to non-trivial {\em gravitational emptiness\/} since, as
can be vividly seen from the integrand in the r.h.s. of (\ref{1093a}),
they imply the vanishing of the energy-momentum current,
$\Sigma^{\gamma}=0$, hence reducing eq. (\ref{1094}) to the empty
space equation,
$R_{\alpha\beta}\wedge\star\of{\vartheta^{\alpha}\wedge 
\vartheta^{\beta}\wedge\vartheta^{\gamma}}=0$. We therefore conclude
that gauge field configurations which are ``self-dual'' solutions of the
YM vacuum equation exert no influence on the ({\em Riemannian\/})
spacetime in which they dwell.\Par   
Finally, varying $\LL_{LC}\of{\vartheta,\omega,A}$ with respect to the 
Riemannian connection $\omega$ reproduces the Riemannian relation 
$\of{D_{\omega}\vartheta}^{\alpha}=0$ (see \cite[eqs. (2.2)-(2.4)]{G}  
%\footnote{The 2-form $F$ does not depend on $\omega$ for the same 
%reason it does not depend on the choice of a coframe.} 
for the detailed derivation) and therefore provides us with no further
information. Hence eqs. (\ref{1094}) and (\ref{1099}) alone capture
the entire dynamics of the $\of{Y_{n}\off{A},g}$ merger.
%--------------------------
\subsection{\sf The Post-Riemannian action in a fixed
non-metric gauge}\label{s43} 
%-------------------------  
In the more general case, $\gb{\omega}{\varphi}\neq0$, and 
$\R_{\off{\alpha\beta}}$ doesn't necessarily depend on $\omega$
solely. Using formula (\ref{620}), and after making some
rearrangements, we arrive at the following expression for the
Lagrangian at the gravitational
sector:\footnote{$\R_{\alpha\beta}\of{\varpi}\wedge\star 
\of{\vartheta^{\alpha}\wedge\vartheta^{\beta}}$ 
contains also the term
$A\wedge{A}\otimes\Q^{\gamma}_{\;\:\beta}\Q_{\alpha\gamma}$ but its 
capital trace vanishes identically.}
\beq{1100}
\LL_{GS}&=&
\Trr{R_{\alpha\beta}\of{\omega}\otimes\bs{1}_{\rho\of{U}}+
\gb{\omega}{\varphi}_{\alpha\beta}+R_{\alpha\beta}\of{\varphi}}
\wedge\star\of{\vartheta^{\alpha}\wedge\vartheta^{\beta}}\nonumber\\&=&
\Trr{R_{\alpha\beta}\of{\omega}\otimes\bs{1}_{\rho\of{U}}-
A\wedge\of{D_{\omega}\Q}_{\alpha\beta}+dA\otimes\Q_{\alpha\beta}}
\wedge\star\of{\vartheta^{\alpha}\wedge\vartheta^{\beta}}.\nonumber\\
\eeq
Clearly, if $A$ takes values in a traceless algebra, the  
$A$-dependent terms disappear, and the action for the gravitational
sector reduces to the ordinary EH action of GR. Otherwise, if $A$
carries a trace, the gravitational Lagrangian contains new mixing
terms in which the YM potential couples to the gravitational
field.\Par  
In order to make the picture in this genuine case more
transparent we pick a direction for $\bs{\qq}_{\alpha}$, and
rewrite the action in that gauge. This can be consistently done if we
assume that it is possible to foliate each spacetime
patch into slices of $\of{n-1}$-hypersurfaces. We may then align
$\bs{\qq}_{\alpha}$, in each of these patches, in the direction normal 
to the hypersurfaces, say along $\bs{e}_{0}$, and assign:
\be{1114}
\bs{\qq}_{\alpha}=:\frac{1}{n}Y\of{x}\delta_{\alpha}^{0}\bs{e}_{0},
\ee
with an arbitrary {\em scaling function\/}
$Y\of{x}$.\footnote{\,$Y$ can be interpreted as a local mussure
for the lack of metricity; a vanishing $Y$ corresponds to metricity.}
Note that by making this choice we haven't fallen back into a
$\of{Y_{n}\off{A},g}$ because now
$\bs{\qq}_{\alpha}\neq\bs{e}_{\alpha}$.\Par 
With this alignment, and from eqs. (\ref{530}), (\ref{540}), and
(\ref{542}), the components of the distortion, the non-metricity, and
the torsion read:  
\beq{1116}
Q_{\;\:\alpha}^{\beta}&\!=\!&
-\frac{1}{n}A\otimes{Y}\delta^{\beta}_{0}\delta_{\alpha}^{0}\,,\\
Q_{\of{\alpha\beta}}&\!=\!&
-\frac{1}{n}A\otimes{Y}\of{\delta_{\alpha}^{0}
g_{0\beta}+\delta_{\beta}^{0}g_{0\alpha}},\label{1116a}\\
T^{\alpha}&\!=\!&
-\frac{1}{n}\vartheta^{0}\wedge{A}\otimes{Y}\delta^{\alpha}_{0}\,,
\label{1117}
\eeq
or, in a matrix form, and in terms of the {\em scale-dependent\/}
effective gauge field $A_{Y}:=A{Y}$,
\beq{1118}
Q_{\of{}}=-\frac{1}{n}A_{Y}\otimes
\of{\ba{c|ccc}2g_{00}&g_{01}&\cdots&g_{0n}\\\hline
g_{10}&&&\\
\vdots&&\mbox{\Large 0}&\\
g_{n0}&&&\ea},
&&
T=-\frac{1}{n}\vartheta^{0}\wedge{A}_{Y}
\otimes\of{\ba{c}1\\0\\\vdots\\0\ea}.\nonumber\\
\eeq
Then, the segmental curvature (eq. (\ref{820})) acquires a particularly
simple structure: 
\beq{1120}
sc\of{\varphi}&=&d\of{A\otimes{\Q}_{\;\:\alpha}^{\alpha}}+
\of{A\otimes{\Q}_{\;\:\alpha}^{\gamma}}\wedge
\of{A\otimes{\Q}_{\;\:\gamma}^{\alpha}}
\nonumber\\&=&
\frac{1}{n}
\of{d{A_{Y}}+\frac{1}{n}{A_{Y}}\wedge{A_{Y}}}
\;=:\;\frac{1}{n}F_{1/n}\of{A_{Y}},
\eeq
namely, it is simply the YM curvature for the scale-dependent gauge
field $A_{Y}$ (with a meaningless prefactor $1/n$). From
eqs. (\ref{700}) and (\ref{1116}), under the gauge, 
\be{1128}
\ba{rcl}
A_{Y}\otimes\delta_{\alpha}^{0}\delta_{0}^{\beta}
&\mapsto&
uA_{Y}u^{-1}\otimes\delta_{\alpha}^{0}\delta_{0}^{\beta}
+nudu^{-1}\otimes\delta_{\alpha}^{\beta}\,;
\ea
\ee
therefore, that single element in $Q_{\;\:\alpha}^{\beta}$ which is
not a pure gauge,\footnote{According to (\ref{1128}) the null entries
in $\mbox{diag}\of{\varphi}$ are indeed null only up to a pure gauge,
$A_{PG}\sim{n}udu^{-1}$.} namely its $00$-component given by $A_{Y}$,
transforms as:   
\be{1130}
\ba{rclcrcl}
A_{Y}&\mapsto&uA_{Y}u^{-1}+nudu^{-1},
&\mbox{whence}&F_{1/n}\of{A_{Y}}
&\mapsto&uF_{1/n}\of{A_{Y}}u^{-1}.
\ea
\ee\Par
Written in the alignment (\ref{1114}) the extended EH action
$S_{LC}\of{\vartheta,\omega,\varphi}$ given by (\ref{1010}) takes 
the compelling form:
\beq{1140}
S_{LC}^{gf}\of{\vartheta,\omega,A_{Y}}&\!=\!&
\frac{\dim\rho}{\ell^{n-2}}\int_{\Omega} 
R_{\alpha\beta}\of{\omega}\wedge\star
\of{\vartheta^{\alpha}\wedge\vartheta^{\beta}}\;+\;
\frac{1}{\ell^{n-4}n^{2}}\int_{\Omega}
\Trr{F_{1/n}\of{A_{Y}}\wedge\star{F}_{1/n}\of{A_{Y}}}
\nonumber\\&\!\!&+\;
\frac{1}{n\ell^{n-2}}\int_{\Omega}\Trr{F_{1/n}\of{A_{Y}}}
\wedge\star\of{\vartheta_{0}\wedge\vartheta^{0}}
\nonumber\\&\!\!&+\;\frac{1}{n\ell^{n-2}}\int_{\Omega}\Trr{
A_{Y}}\wedge\off{\omega_{\alpha0}
\wedge\star\of{\vartheta^{\alpha}\wedge\vartheta^{0}}
-\omega^{0\alpha}\wedge\star\of{\vartheta_{0}\wedge\vartheta_{\alpha}}}.
\eeq
Note that the last term in (\ref{1140}) is not preserved by internal
gauge transformations, and it actually describes an interaction
written in a 
fixed internal gauge. In fact, as one may directly infer from the
detailed derivation of (\ref{1140}), aligning the deformation element
$\bs{q}_{\alpha}$ in a certain direction (which is the same as to
fix the direction of the torsion and non-metricity), amounts to
{\em fixing a gauge\/} also in the internal space for those cases where 
$A$ carries a trace.\footnote{Hence the superscript ``$gf$'' in
$S_{LC}^{gf}\of{\vartheta,\omega,A_{Y}}$.}\Par 
If $A_{Y}$ and $F_{1/n}\of{A_{Y}}$ are traceless objects (in which
case the symmetry on the frames is restricted to local
pseudo-rotations, see section \ref{s35}), the two pieces in
$S_{LC}\of{\vartheta,\omega,A_{Y}}$ with the prefactor $1/n\ell^{n-2}$
in front drop out. Then the gauge invariance of the action is 
retained, and the internal world, at least at the level of the action,
decouples from the gravitational world. The corresponding e.o.m.'s are
those given by eqs. (\ref{1094}) and (\ref{1099}), with $A$ being
replaced by the scale-dependent field $A_{Y}$, $F\of{A}$ being
replaced by $F_{1/n}\of{A_{Y}}$, and a different numerical prefactor,
$\kappa'=\kappa/n^{4}$.\Par    
Otherwise, if $A_{Y}$ has a non-vanishing trace, its trace part (or
its ``photonic'' component) interacts with the gravitational field
$\omega$ via the contact terms $\Tr{A_{Y}}\wedge\omega$. This
interaction, as well as the `vortex' term $\Tr{F_{1/n}}$, are of weak
magnitude, namely $1/\of{n\,\mbox{dim}\rho}$ times the gravitational 
magnitude. It is interesting to note that under the restriction to
orthonormal frames, the two contact terms, and the vortex term,
extinguish even for an {\em extrinsic\/} YM potential that has been
placed by hand: in this case, since we have already been employing
an-holonomic basis, $g^{\alpha\beta}=\eta^{\alpha\beta}$,
$\eta^{\alpha0}=\eta_{\alpha0}=\pm\delta_{\alpha0}$, whence 
\be{1132}
\ba{rcl}
\eta_{\alpha0}\vartheta^{\alpha}\wedge\vartheta^{0}
\,=\,
\pm\vartheta^{0}\wedge\vartheta^{0}\;=\;0,&\mbox{and}&
\omega^{0\alpha}\wedge\star\of{\vartheta_{0}\wedge\vartheta_{\alpha}}
\,=\,
\omega_{\alpha0}\wedge\star\of{\vartheta^{\alpha}\wedge\vartheta^{0}},
\ea
\ee
so that the terms with the prefactor ${1/n\ell^{n-2}}$ in
(\ref{1140}) drop-out. Hence the new interactions cannot be
detected in a $V_{n}$ whatsoever.\Par  
It is admitted that the dynamics of the scaling function $Y$ has
been completely absorbed in the dynamics of the redefined potential
$A_{Y}$. But: as $Y\of{x}$ 
encodes the microstructure of spacetime, different spacetime scales
display different gluon structures, and this, in turn, may affect
the phenomenology. For example, in the case of a traceless $A$,
and for extremely small values of $Y$
(compared to $A$), or when $Y$ heavily fluctuates such that
$Y^{2}$ becomes negligible compared to $dY$, we have:
$F_{1/n}\of{A_{Y}}\approx{d}A_{Y}$, whence
\be{1134}
L_{YM}\approx\Tr{dA_{Y}\wedge\star\,{d}A_{Y}},
\ee
and the internal (non-Abelian) world becomes effectively free. 
%---------------
\section{\sf Summary and closing remarks}\label{s5}
%---------------
The local texture of the frame bundle as a space product between the
tangent frame and the base enables the {\em merging\/} of 
gravity with YM theory under the roof of post-Riemannian
geometry. The spacetime geometry, which stands on three structural
identities, and on the gauge symmetry induced on the frames, is not
affected by internal gauge transformations, despite that the
structural identities in this framework are loaded with internal
degrees of freedom; furthermore, its metric part remains real-valued
even for complex-valued internal symmetry groups. A post-Riemannian
spacetime endowed with an internal symmetry was termed {\em merged spacetime\/}
because the corresponding two symmetry structures, frames and
internal, were merged at the gauge sector level into a single
spacetime fabric.\Par   
From a tangent bundle perspective, the internal symmetry lies entirely
on the basespace, whereas gravity emerges as the local symmetry
structure associated with the frames. In this sense, a merged
spacetime is a bundle setup where the fiber {\em and\/} the base play
an active role. Algebraically speaking, excluding the Weyl-Cartan case, 
the overall gauge structure is not a Whitney product of two
vector bundles since the generating algebra is not a simple sum of
algebras. From the physical perspective, {\em the torsion and non-metricity
in spacetime, and the internal-space potential, are one and the
same thing in their basespace components\/}.\Par
The merging of gravity with YM interactions could not have been
correctly established without referring explicitly 
to the post-Riemannian extension of GR. Had we originally split
the Riemannian connection into its base-part and fiber-part, and
attributed internal degrees of freedom to the former, we would have
necessarily contaminated the frames, the coframes, and the metric
tensor (which we had to symmetrize a-priori) by these degrees of
freedom. In this case, not only we would have lost commutativity
(the frames and coframes would have become non-Abelian), but 
the gravitational sector and the internal symmetry sector could not
have been separated and split apart. Furthermore, since 
gravity is real-valued, any symmetry that dwells at the metric 
sector must be real-valued as well.\Par
A motivated (and rather minimal) choice of action, based on an extended
variant of the EH action, led to a natural split between the
gravitational branch and the internal branch of the merger. In the 
Weyl-Cartan scenario, the gravitational branch is seen to consists of
pure Riemannian gravity, while the internal branch displays the 
structure of pure YM, with its field strength building up the
energy momentum current in Einstein's equation. It was shown that
the energy momentum current is traceless only in four
dimensions, and that the star eigenform curvature solutions to the
YM field equations (in four dimensions) lead to an empty space 
Einstein equation. We therefore conclude that a self-dual YM potential
has no dynamical effect on the Riemannian spacetime in which it 
dwells.\Par  
In the more general case, in a fixed distortion gauge, the
gauge fields effectively become scale-dependent, and consequently
sensitive to the non-metric microstructure of spacetime. A glance at
the detailed structure of
the gauge-fixed action reveals that the YM potential weakly interacts
with the gravitational potential via the ``photonic'' components
it carries. Furthermore, the scaling function attributed to
the gluons is seen to determine qualitatively what type of effective
theory shows at the various situations; in particular, an
effectively free theory is anticipated at weak, or at a heavily
fluctuating non-metricity.\Par 
Our formalism obviates the need to introduce a symmetry-breaking
mechanism for the post-Riemannian GR (so as to comply with the
equivalence principle). Such mechanism was advocated in \cite{SN2},
and later elaborated and summarized in \cite{HMMN}. The trigger that
ignites the symmetry-breaking in these studies is a dilaton field
that originates from conformal symmetry added to the post-Riemannian
framework in the manner Weyl 
added conformal symmetry to Einstein's gravity. In this way the gauge
fields that correspond to shears and dilations become massive. In the
merger setup, however, the metric tensor and the line-element
are absolutely closed, and their absolute closure plays a role
similar to the role played by the equivalence principle in a metric
spacetime; symmetry breaking therefore becomes redundant.\Par  
It is impressing that gravity and non-Abelian YM
theory are naturally interlaced into such an elegant fabric. Without
supersymmetry, and without need for extra dimensions, the two
physical sectors coexist as two complementary symmetry structures that
make-up a single spacetime entity - the merger. The ultimate goal in
this direction would probably be to extend the validity of the theory
to the quantum regime, or may be to extract quantum phenomena
from the merger setup itself. This task would probably
require further refinements in the underlying geometrical framework.
Perhaps it even requires giving-up locality in which case the
whole setup should be entirely revised.
\appendix
%----------------------------
\section{Integration formulas and topological considerations}\label{a} 
%----------------------------
%----------------------
\subsection{\sf The 1-st Chern class and its associated
charge}\label{a1}  
%----------------------
The 1-st Chern class associated with the frame bundle can
easily be calculated using the 3-rd structural identity, 
$\of{D_{\varpi}Q}_{\of{\alpha\beta}}=-R_{\of{\alpha\beta}}$. 
Transvecting $\of{D_{\varpi}Q}_{\of{\alpha\beta}}$ with
$g^{\alpha\beta}$, and `smuggling' the latter into the exterior
derivative of $Q_{\of{\alpha\beta}}$ gives: 
\be{412}
g^{\alpha\beta}\of{D_{\varpi}Q}_{\of{\alpha\beta}}=
d\of{g^{\alpha\beta}Q_{\of{\alpha\beta}}}.
\ee
Therefore, the trace of the 3-rd identity reads
\be{416}
R^{\alpha}_{\;\:\alpha}=
d\varpi^{\alpha}_{\;\:\alpha}=
-dQ^{\alpha}_{\;\:\alpha}.
\ee\Par
Let us now construct a $GL\of{n,\mathbb{R}}$ frame-bundle, whose basespace
has the topology of a 2-sphere $S^{2}$. The Weyl covector
$Q^{\alpha}_{\;\:\alpha}$ on the northern and southern hemispheres is
denoted, respectively, by 
${}^{+}Q^{\alpha}_{\;\:\alpha}$ and ${}^{-}Q^{\alpha}_{\;\:\alpha}$.
Integrating $R^{\alpha}_{\;\:\alpha}$ on the sphere, making use of
eq. (\ref{416}), and employing Stokes' theorem gives: 
\be{417}
\int_{S^{2}}R^{\alpha}_{\;\:\alpha}\,=\,
-\,\oint_{S^{1}}{}^{+}Q^{\alpha}_{\;\:\alpha}
\,+\,\oint_{S^{1}}{}^{-}Q^{\alpha}_{\;\:\alpha}
\,=\,0,
\ee
because $Q^{\alpha}_{\;\:\alpha}$ is gauge-invariant over the whole
sphere (hence
${}^{+}Q^{\alpha}_{\;\:\alpha}={}^{-}Q^{\alpha}_{\;\:\alpha}$). In
the frame geometrical picture
$\varpi^{\alpha}_{\;\:\alpha}$ ($=-Q^{\alpha}_{\;\:\alpha}$) is  
gauge invariant since the non-linear terms generated in the
transformation of $\varpi^{\beta}_{\;\:\alpha}$ come solely from its
rotational piece.\Par  
Consider next a $GL\of{n,\mathbb{R}}\times{U}\of{N,\mathbb{C}}$ {\em 
foliar\/} bundle \cite{M1} whose basespace is topologically a
2-sphere. In this case, as we have already seen, the segmental
curvature takes the form: 
\be{460}
\R^{\alpha}_{\;\:\alpha}\of{\varpi}
=d\varphi^{\alpha}_{\;\:\alpha}+
\varphi^{\alpha}_{\;\:\beta}\wedge\varphi^{\beta}_{\;\:\alpha}.
\ee
We denote the distortion 1-form on the northern and southern
hemispheres by ${}^{+}\varphi$ and ${}^{-}\varphi$, respectively.
Let $k_{+-}$ be the transition function from the northern hemisphere
to the southern hemisphere for the $U\of{N,\mathbb{C}}$ sub-bundle,
and let also $h_{+-}$ be the corresponding transition function for the
$GL\of{n,\mathbb{R}}$ sub-bundle. Making use of formula (\ref{700}),
the trace in frame-space (tr) of the distortion 1-form on the two
hemispheres satisfies: 
\beq{462}
\tr{{}^{+}\varphi}&=&
k_{+-}\trr{h_{+-}\of{^{-}
\varphi}h_{+-}^{-1}}k_{+-}^{-1}+
nk_{+-}dk_{+-}^{-1}\nonumber\\
&=&k_{+-}\tr{^{-}\varphi}k_{+-}^{-1}+
nk_{+-}dk_{+-}^{-1};
\eeq
taking the trace of (\ref{462}) also in the representation space of
$U$ (Tr) yields,
\be{464}
\mbox{Tr\,}\tr{{}^{+}\varphi}=\mbox{Tr\,}\tr{{}^{-}\varphi}+
n\Tr{k_{+-}dk_{+-}^{-1}}.
\ee
Note that the term $\tr{h_{+-}dh_{+-}^{-1}}$ is absent in formula
(\ref{462}) because such a term would have emerge from the rotational
piece in the connections (where the derivatives are found), and not
from the distortion 1-form which transforms covariantly as a
frame-space coordinate tensor.\Par 
Employing Stokes' theorem, and making use of formula (\ref{464}), the
generalized 1-st Chern class associated with the foliar bundle, namely
$\mbox{Tr\,tr\,}{\R}$, is integrated to yield:\footnote{Note that
$\mbox{Tr\,}\tr{\varphi\wedge\varphi}$ identically vanishes due to the
fact that $\varphi$ is a 1-form.}
\be{466}
\int_{S^{2}}\mbox{Tr\,tr\,}{\R}=
n\oint_{S^{1}}\Tr{k_{+-}dk_{+-}^{-1}}.
\ee
If the internal gauge field $A$ is Abelian, the charge simply equals
$2\pi{mn}$, $m\in\mathbb{Z}$ being the winding number associated with
$S^{1}$. Otherwise, if $A$ is non-Abelian, the charge 
is characterized by the mappings of the circle into domains in 
group-space,\footnote{$udu^{-1}=u\off{u^{-1}\of{\theta+d\theta}- 
u^{-1}\of{\theta}}\in{U}$, where $\theta\in\uu$ is the
non-Abelian group angle.} and its magnitude is determined by the
volume of these domains. For a traceless $A$, the charge vanishes
whatsoever.\Par  
%------------------
\subsection{\sf Non-Abelian Stokes' theorem for the Weyl
covector}\label{a12} 
%------------------
Integrating formula (\ref{412}) 
over a compact 2-domain $\Sigma$, whose boundary is
$\partial\Sigma$, assuming that $g^{\alpha\beta}Q_{\of{\alpha\beta}}$ is 
nowhere singular there, yields 
\be{470}
\int_{\Sigma}g^{\alpha\beta}\of{D_{\varpi}Q}_{\of{\alpha\beta}}
=\oint_{\partial\Sigma}g^{\alpha\beta}Q_{\of{\alpha\beta}},
\ee
which may possibly be interpreted as the non-Abelian analogue of Stokes'
theorem for the non-metricity-trace (or Weyl's covector).\Par  
In an $\of{L_{n}\off{A},g}$, however, $Q_{\of{\alpha\beta}}$ no longer
commutes with itself. Consequently, formula (\ref{412}) should be
replaced by
\be{474}
g^{\alpha\beta}\of{D_{\varpi}Q}_{\of{\alpha\beta}}=
d\of{g^{\alpha\beta}Q_{\of{\alpha\beta}}}+ 
Q^{\of{\alpha\beta}}\wedge{Q}_{\of{\alpha\beta}}.
\ee
Having in mind that the non-Abelian $Q_{\of{\alpha\beta}}$ transforms
non-linearly with respect to internal gauge transformations, we
should better select a 2-domain $\Sigma$ that can be covered by a 
{\em single\/} patch over which $Q_{\of{\alpha\beta}}$ is globally
defined. Integration now yields:  
\be{490}
\int_{\Sigma}\off{g^{\alpha\beta}\of{D_{\varpi}Q}_{\of{\alpha\beta}}
-Q^{\of{\alpha\beta}}\wedge{Q}_{\of{\alpha\beta}}}
=\oint_{\partial\Sigma}g^{\alpha\beta}Q_{\of{\alpha\beta}}.
\ee
This has been shown to be equivalent to the {\em Abelian\/} Stokes'
theorem for the internal gauge field $A$ in a $\of{Y_{n}\off{A},g}$,
see eqs. (\ref{1078})-(\ref{1092}), section \ref{s42}.   
%------------------
\subsection{\sf Integration formulas for the transvected
torsion}\label{a13}
%------------------
A kind of Stokes' theorem that involves covariant exterior derivatives
can be derived also for the transvected torsion,
$\vartheta_{\alpha}\wedge{T}^{\alpha}$. This is done with the aid of 
the Nieh-Yan (NY) topological 4-form whose ordinary post-Riemannian
format reads:  
\be{410}
-Q_{\of{\alpha\beta}}\wedge\vartheta^{\alpha}\wedge{T}^{\beta}
-g_{\alpha\beta}\of{T^{\alpha}\wedge{T}^{\beta}+
\vartheta^{\alpha}\wedge\vartheta^{\gamma}\wedge{R}_{\;\:\gamma}^{\beta}}
\;=\;d\of{\vartheta_{\alpha}\wedge{T}^{\alpha}}\label{405};
\ee
note that the NY density is a transvection field - an invariant. For
recent analytical accounts of the various NY densities, the reader is
referred to \cite{CZ,GWZ}, and \cite{OMBH}.\Par 
Taking the integral of eq. (\ref{410}) over a simply-connected
4-domain enclosed by a boundary that can be covered by a single patch 
(for example, the 4-ball) yields: 
\be{418}
\int_{\Sigma}
\off{-Q_{\of{\alpha\beta}}\wedge\vartheta^{\alpha}\wedge{T}^{\beta}
-g_{\alpha\beta}\of{T^{\alpha}\wedge{T}^{\beta}+
\vartheta^{\alpha}\wedge\vartheta^{\gamma}\wedge{R}_{\;\:\gamma}^{\beta}}}
=\oint_{\partial\Sigma}\vartheta_{\alpha}\wedge{T}^{\alpha}.
\ee
From the deformation criterion, and from the 1-st structural identity
we have:
\be{420}
\ba{rcl}
-Q_{\of{\alpha\beta}}\wedge\vartheta^{\alpha}\wedge{T}^{\beta}&=&
-Q_{\alpha\beta}\wedge\vartheta^{\alpha}\wedge{T}^{\beta}+
T_{\alpha}\wedge{T}^{\alpha},\\
\mbox{and}\;\;\:
\vartheta^{\alpha}\wedge\vartheta^{\beta}\wedge{R}_{\alpha\beta}&=&
\vartheta_{\alpha}
\wedge\of{D_{\varpi}T}^{\alpha},
\ea
\ee
respectively. With the aid of these two relations, formula
(\ref{418}) can be rewritten in the compelling form: 
\be{421}
\int_{\Sigma}
\vartheta_{\alpha}\wedge\off{T^{\beta}
\wedge{Q}^{\alpha}_{\;\:\beta}-\of{D_{\varpi}T}^{\alpha}}
=\oint_{\partial\Sigma}
\vartheta_{\alpha}\wedge{T}^{\alpha},
\ee
which one may tentatively interprets as the non-Abelian generalization
of Stokes' theorem for the transvected torsion
$\vartheta_{\alpha}\wedge{T}^{\alpha}$.\footnote{In a Weyl-Cartan
spacetime, the two integrands in eq. (\ref{421}) identically
vanish.}\Par 
Otherwise, consider a $GL\of{n,\mathbb{R}}$ frame-bundle whose
basespace is topologically a 4-sphere $S^{4}$. In this case the
integration of the r.h.s of eq. (\ref{410}), which splits into two
integrations over the boundaries of the northern and southern
4-hemispheres (namely, over the equatorial 3-sphere), vanishes due to
the fact that $\vartheta_{\alpha}\wedge{T}^{\alpha}$ is everywhere
invariant.\footnote{A similar statement was already made in \cite{GWZ}.}  
Making use of the upper equation in (\ref{420}) we arrive at the
relation: 
\be{426}
\int_{S^{4}}
\vartheta^{\alpha}\wedge\vartheta^{\beta}\wedge{R}_{\alpha\beta}\,
=\int_{S^{4}}
\vartheta^{\alpha}\wedge{T}^{\beta}\wedge{Q}_{\alpha\beta}.
\ee\Par
We now turn to the $\of{L_{n}\off{A},g}$ case. Here, due to the
non-commutativity of the distortion $\varphi$, one encounters
an additional commutator term in formula (\ref{410}):
\beq{480}
-Q_{\of{\alpha\beta}}\wedge\vartheta^{\alpha}\wedge{T}^{\beta}
-g_{\alpha\beta}\of{T^{\alpha}\wedge{T}^{\beta}+
\vartheta^{\alpha}\wedge\vartheta^{\gamma}\wedge{\R}_{\;\:\gamma}^{\beta}}
-\vartheta_{\alpha}\wedge\off{\!\!\off{Q^{\alpha}_{\;\:\beta},T^{\beta}}\!\!}
&=&d\of{\vartheta_{\alpha}\wedge{T}^{\alpha}};\nonumber\\
\eeq
A compelling property of the torsion in a merged spacetime is
that, despite the fact that $T^{\alpha}$ transforms 
non-linearly with respect to internal gauge transformations, 
$\vartheta_{\alpha}\wedge{T}^{\alpha}$ transforms covariantly
after all.\footnote{This property is shared also by the off-diagonal
elements of $\varphi_{\;\:\alpha}^{\beta}$.}
This follows directly from formula (\ref{770}):  
\be{482}
\ba{rclcrcl}
T^{\alpha}&\mapsto&{u}T^{\alpha}u^{-1}+udu^{-1}\wedge\vartheta^{\alpha}
&\Rightarrow&
\vartheta_{\alpha}\wedge{T}^{\alpha}&\mapsto&
\vartheta_{\alpha}\wedge{u}{T}^{\alpha}u^{-1}.
\ea
\ee\Par
%since $\vartheta_{\alpha}\wedge\vartheta^{\alpha}\equiv0$.\Par
Consider then a $GL\of{n,\mathbb{R}}\times{U}\of{N,\mathbb{C}}$
{\em foliar\/} bundle whose basespace is taken to be the 4-sphere. 
Integrating the trace (Tr) of formula (\ref{480}) over the sphere,
making use of the fact that
$\Tr{\vartheta_{\alpha}\wedge{T}^{\alpha}}$ is an invariant of the two
symmetries in the foil, we get: 
\beq{484}
\int_{S^{4}}
\Tr{\vartheta^{\alpha}\wedge\vartheta^{\beta}\wedge{\R}_{\alpha\beta}}
&=&\int_{S^{4}}
\Tr{-Q_{\alpha\beta}\wedge\vartheta^{\alpha}\wedge{T}^{\beta}
-\vartheta_{\alpha}\wedge\bgb{Q^{\alpha}_{\;\:\beta}}{T^{\beta}}}
\nonumber\\&=&
\int_{S^{4}}
\Tr{\vartheta^{\alpha}\wedge{T}^{\beta}\wedge{Q}_{\alpha\beta}};
\eeq
Hence, formula (\ref{484}) generalizes formula (\ref{426}) to the case
of $\of{L_{n}\off{A},g}$.\footnote{Making use of the 
transformation rules for $Q_{\alpha\beta}$ (eq. (\ref{710})) and
$T^{\alpha}$ (eq. (\ref{770})), the reader may verify that 
$\Tr{\vartheta^{\alpha}\wedge{T}^{\beta}\wedge{Q}_{\alpha\beta}}$ 
is indeed (internally) gauge-invariant, as eq. (\ref{484})
requires.} Finally, one may easily verify that the non-Abelian
extension of Stokes' theorem for the transvected torsion in an
$\of{L_{n},g}$, as given by eq. (\ref{421}), holds in this form also
in an $\of{L_{n}\off{A},g}$. 
\newpage
%---------------------------
\vspace{0.5cm}\Par\noindent
{\large\bf\sf Acknowledgments}\Par\Par\noindent
I am indebted to Yuval Ne'eman, my supervisor during my Ph.D. studies,
for many fruitful conversations, for helpful suggestions, and for
providing a financial support for this work. 
I also thank Friedrich W. Hehl for warm hospitality during my short
visit to $\ddot{\mbox{C}}$oln in June '98, and for referring me to the
paper by  P. Teyssandier \& R. W. Tucker on Gravity, Gauges and Clocks
\cite{TT}. Gratitude to Larry Horowitz and Richard Kerner for their
valuable comments on the manuscript.  
%---------------------------%---------------------------
%===========================

\end{document}